\documentclass[prb,aps,twocolumn,reprint,amsmath,amssymb,showpacs,superscriptaddress]{revtex4-1}

\usepackage{verbatim}

\usepackage{txfonts}
\usepackage{bm}
\usepackage{dcolumn}
\usepackage{graphicx,tikz}

\usepackage{amsfonts}
\usepackage{amsmath}
\usepackage{amssymb}
\usepackage{subfigure}

\usepackage{color}

\usepackage{times}

\usepackage[colorlinks,bookmarks=false,citecolor=blue,linkcolor=blue,urlcolor=blue,pdfstartview=FitH]{hyperref}

\newcommand{\be}{\begin{equation}}
\newcommand{\ee}{\end{equation}}
\newcommand{\bea}{\begin{eqnarray}}
\newcommand{\eea}{\end{eqnarray}}
\newcommand{\fmtilde}{$\widetilde{\text{FM}}$ }

\newcolumntype{.}{D{.}{.}{-1}}

\def\bs{\boldsymbol}
\def\vec{\mathbf}
\def\mc{\mathcal}

\allowdisplaybreaks

\begin{document}

\title{Kitaev anisotropy induces mesoscopic $Z_2$ vortex crystals in frustrated hexagonal antiferromagnets}

\author{Ioannis Rousochatzakis}
\affiliation{Institute for Theoretical Solid State Physics, IFW Dresden, Helmholtzstr.~20, 01069 Dresden, Germany}
\affiliation{School of Physics and Astronomy, University of Minnesota, Minneapolis, MN 55455, USA}

\author{Ulrich K. R\"ossler}
\affiliation{Institute for Theoretical Solid State Physics, IFW Dresden, Helmholtzstr.~20, 01069 Dresden, Germany}

\author{Jeroen \surname{van den Brink}}
\affiliation{Institute for Theoretical Solid State Physics, IFW Dresden, Helmholtzstr.~20, 01069 Dresden, Germany}

\author{Maria Daghofer}
\affiliation{Institute for Theoretical Solid State Physics, IFW Dresden, Helmholtzstr.~20, 01069 Dresden, Germany}
 \affiliation{Institute for Functional Materials and Quantum
   Technologies, University of Stuttgart, Pfaffenwaldring 57
D-70550 Stuttgart, Germany}

\date{\today}

\begin{abstract}
The triangular-lattice Heisenberg antiferromagnet (HAF) is known to carry topological $Z_2$ vortex excitations which form a gas at finite temperatures. Here we show that the spin-orbit interaction, introduced via a Kitaev term in the exchange Hamiltonian, condenses these vortices into a triangular $Z_2$ vortex crystal at zero temperature. The cores of the $Z_2$ vortices show abrupt, soliton-like magnetization modulations and arise by a special intertwining of three honeycomb superstructures of ferromagnetic domains, one for each of the three sublattices of the 120$^\circ$ state of the pure HAF.  This is a new example of a nucleation transition, analogous to the spontaneous formation of magnetic domains, Abrikosov vortices in type-II syperconductors, blue phases in cholesteric liquid crystals, and skyrmions in chiral helimagnets. As the mechanism relies on the interplay of geometric frustration and spin-orbital anisotropies, such vortex mesophases can materialize as a ground-state property in spin-orbit coupled correlated systems with nearly hexagonal topology, as in triangular or strongly frustrated honeycomb iridates. 
\end{abstract}
\maketitle

\section{Introduction}
Topological defects in an ordered state cannot be removed by small modifications of the underlying system, making them rather stable even if they cost energy. Due to this stability and their localized character, they behave in many respects like ``particles'' whose quantum-numbers and properties are determined by the host system. They are often quite exotic, as for example defects in strongly frustrated ``spin-ice'' compounds, which behave like magnetic monopoles~\cite{Castelnovo:2008hb,Bramwell:2009ep}. Other topological defects are domain walls with Yukawa-like interactions~\cite{Yukawa} or vortices in manganites~\cite{PhysRevX.2.041022}. A particular kind of defects, $Z_2$ vortices, are their own anti-particles: two $Z_2$ vortices can annihilate each other. They can be thermally excited in host systems with an SO(3) order parameter like the superfluid A-phase of $^3$He~\cite{refId0,Anderson:1977de,Mermin1979}, spinor Bose-Einstein condensates~\cite{SpinorBEC}, and the triangular-lattice Heisenberg antiferromagnet (HAF)~\cite{Kawamura:1984hk}. 

Apart from defining topological excitations, particle-like modulations can also condense into a lattice at thermodynamic equilibrium~\cite{DesGennes1975}. Stable localized solutions to classical field theories were first introduced by Skyrme~\cite{Skyrme1961} in order to explain how discrete particles can arise out of a continuum field background. However, the classical theorem by Hobart\cite{Hobart1963} and Derrick\cite{Derrick1964} poses severe restrictions on the type of non-linear classical field theories that stabilizes `particles'. A standard mechanism to evade these restrictions is operative in condensed-matter systems with a fixed `handedness', i.e., systems without inversion symmetry~\cite{Bogdanov1995}. In the long-wavelength limit, the handedness manifests in the form of Lifshitz invariants (linear gradient terms)~\cite{Dz64} which favor a twisting of the order parameter along more than one spatial direction, thus allowing for localized modulations. This universal mechanism underlies the condensed-matter examples of Abrikosov vortices in type-II superconductors~\cite{Abrikosov1957}, ``double-twist tubes'' in blue phases of cholesteric liquid crystals~\cite{Wright89}, and skyrmions in non-centrosymmetric helimagnets~\cite{Bogdanov1989,Roessler2006,muehlbauer2009,tonomura2012,yu2012,yu2010,Seki2012,oleg2014}.

\begin{figure}[!b]
\includegraphics[width=\columnwidth,angle=0]{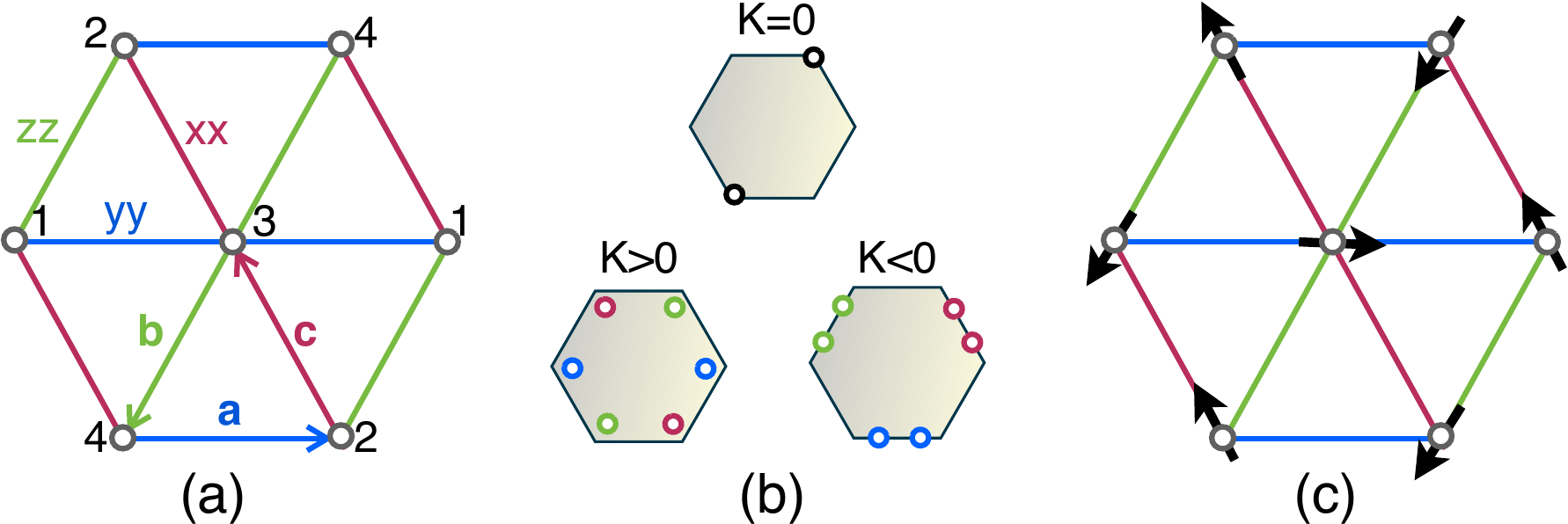}
\caption{(a) Triangular Heisenberg antiferromagnet with additional, Ising-like ``xx'', ``yy'' and ``zz'' interactions, along the three directions of the lattice, $\vec{c}\!=\!-\vec{a}\!-\!\vec{b}$, $\vec{a}\!=\!a\vec{x}'$, and $\vec{b}\!=\!-a(\vec{x}'\!+\!\sqrt{3}\vec{y}')/2$ ($a$ is the lattice constant), respectively, see Eq.~(\ref{eq:ham}). Here $\{\vec{x}',\vec{y}'\}$ define the plane of the lattice and are fixed by the spin-orbit coupling, see Sec.~\ref{sec:model}. (b) The positions $\vec{Q}^{(\gamma)}$ [$\gamma\!=\!x$ (red), $y$ (blue), $z$ (green)] of the minima of the three coupling functions $\lambda_\gamma(\vec{k})$ of Eq.~(\ref{eq:lambda}) inside the first Brillouin zone (BZ), for small (positive or negative) anisotropy $K$. The $K\!=\!0$ minima at the corners of the BZ are also shown for comparison. (c) The 120$^\circ$ state of the pure Heisenberg model ($K\!=\!0$) in real space.\label{fig:model}} 
\end{figure}

Here, we show that the basic ingredients for the creation of extended phases composed of particle-like modulations -- the fixed `handedness' and the presence of Lifshitz invariants along several spatial directions -- are generically present in frustrated spin-orbit coupled Mott insulators with hexagonal symmetry and 90$^\circ$ bond angles, where the spin-orbit coupling manifests in the form of anisotropic, spatially dependent Ising-like interactions (termed `Kitaev' interactions in the literature)~\cite{Jackeli:2009p2016,Chaloupka:2013ju,Chaloupka:2010p2461}.

We will consider here in more detail the simplest case, the triangular antiferromagnet (AF), whose ground state at the isotropic Heisenberg point is the well-known three-sublattice configuration of Fig.~\ref{fig:model}(c)~\cite{PhysRev.87.290}. Its order parameter is that of a rigid rotator, i.e. SO(3). This state breaks the inversion symmetry spontaneously and so the fixed `handedness' is guaranteed even in the presence of spin-orbit coupling, while the crucial requirement of more than one Lifshitz invariants is fulfilled by the special structure of the Kitaev interactions, as shown by an explicit derivation of the long-distance action of the problem. Our results show that this mechanism stabilizes a triangular superlattice of $Z_2$ vortices with a lattice constant that goes to infinity as we approach the pure Heisenberg limit. 

% The main qualitative features of this $Z_2$ vortex crystal ($Z_2$VC)
% ground state are the following. First, the $Z_2$ vortices arise by a
% special arrangement of three FM domain honeycomb superstructures, one
% for each of the three sublattices of the neighboring 120$^\circ$
% state. The three superstructures are intertwined in such a way that
% the center of a domain in one sublattice (say A) coincides with
% vertices in the superstructures of the other two sublattices (B and
% C). With three domains meeting at each vertex, it follows that the
% spin plane formed by the spins in B and C completes a $2\pi$ rotation
% as we go around the center of the A domain. So the center of each
% sublattice FM domain is the core of a $Z_2$ vortex. 

The main qualitative features of this $Z_2$ vortex crystal ($Z_2$VC) are the following (see Sec.~\ref{sec:Z2}). First, the $Z_2$VC state preserves the threefold rotation symmetry of the model, with spins at the vortex cores pointing along the $\langle111\rangle$ axes. As a result, the three components of the spin structure factor have equal weight, and the corresponding harmonics are related to each other by threefold rotations.  

Second, the cores of the $Z_2$ vortices are defects of the 120$^\circ$ state of the pure Heisenberg limit, with a finite FM canting and a reduced chirality, see Secs.~\ref{sec:Z2nature} and~\ref{sec:cores}. Abrupt soliton-like modulations of the magnetization, see Secs.~\ref{sec:soliton}, suggest that the $Z_2$VC phase arises from the commensurate 120$^\circ$ state through a new example of a nucleation transition.~\cite{DesGennes1975} 

Third, the cores optimize, on the other hand, the energy gain from the Kitaev anisotropy, see Sec.~\ref{sec:particles}. In contrast to thermally induced defects ($Z_2$ vortices in the HAF model~\cite{Kawamura:1984hk,Kawamura:2010di,Kawamura:2011bq,Tamura12}, or $\mathbb{Z}$ vortices driving Berezinsky-Kosterlitz-Thouless transitions~\cite{Berezinski1971,*Berezinski1972,*KT1973}), the $Z_2$ vortices are here thus favored by energy and not by entropy.  

Fourth, the  $Z_2$VC phase survives in a large region of the classical ground-state phase diagram, with the vortex density increasing (possibly in an Devil's
staircase manner) with the strength of the Kitaev anisotropy, see Sec.~\ref{sec:dv}. Maximum density corresponds to
commensurate vortex crystals with periods of one (two)
lattice spacings for negative (positive) Kitaev anisotropy. These
solutions are special members of the infinitely degenerate
classical ground state manifolds at the two
boundaries of the $Z_2$VC phase. However, quantum and thermal
fluctuation working against non-colinear spin patterns likely stabilize
other phases rather than very dense $Z_2$VCs near the phase boundaries.  

Fifth, the $Z_2$ vortices can be identified  by looking at the three sublattices of the 120$^\circ$ state, as done in Sec.~\ref{sec:domains}: Spins on each sublattice form honeycomb superstructures of FM domains. The three superstructures are overlaid in such a way that the center of a domain in one sublattice (say A) coincides with vertices in the superstructures of the other two sublattices (B and C). With three domains meeting at each vertex, it follows that the spin plane formed by the spins in B and C completes a $2\pi$ rotation as we go around the center of the A domain. So the center of each sublattice FM domain is the core of a $Z_2$ vortex.

The remaining part of the article is organized as follows. We begin in Sec.~\ref{sec:model} with the definition of the model and its symmetries. In Sec.~\ref{sec:gpd} we give the global phase diagram for all different values of the Heisenberg and the Kitaev coupling parameters, as obtained from the standard Luttinger-Tisza (LT) minimization method~\cite{LT, *Bertaut, *Litvin, *Kaplan} and our numerical simulations. For the most part, the LT method delivers the correct classical or families of classical ground states, including the so-called nematic states when the system is governed by Kitaev anisotropies alone. In the remaining regions the LT method does not work, but gives the first important insights as to why the situation near the AF Heisenberg point is very special. Sec.~\ref{sec:Z2} focuses entirely on this particular region, with a detailed analysis of our numerical data. These include data from Monte Carlo simulations as well as from an iterative variational minimization scheme that delivers very low energies and accurate predictions for the vortex distance as a function of the Kitaev anisotropy. The physical mechanism for the condensation of $Z_2$ vortices is then discussed in Sec.~\ref{sec:LDT} by the analysis of the long-distance action of the problem, which delivers the Lifshitz invariants as well as the role of a cross-coupling between the SO(3) rotator and the FM canting degree of freedom. Finally, we conclude with a more general discussion on relevant materials and related models in Sec.~\ref{sec:disc}.

\section{Model}\label{sec:model}
The model we consider here is described by the Kitaev-Heisenberg Hamiltonian 
\begin{align}\label{eq:ham}
\mc{H} = J\sum_{\langle ij\rangle} \vec{S}_i\!\cdot\!\vec{S}_j+ 
K\!\sum_{\bs{\epsilon}=\vec{a},\vec{b},\vec{c}}
\sum_{\langle ij\rangle\parallel \bs{\epsilon}} S_i^{\gamma_{\bs{\epsilon}}} S_j^{\gamma_{\bs{\epsilon}}}~,
\end{align}
where $\langle ij\rangle$ labels nearest neighbor (NN) classical spins of unit length on the triangular lattice, and $\gamma_{\vec{a}}\!=\!y$, $\gamma_{\vec{b}}\!=\!z$ and $\gamma_{\vec{c}}\!=\!x$, see Fig.~\ref{fig:model}(c). The first term $\propto\!J$ denotes the isotropic Heisenberg exchange, as it arises in many correlated materials through superexchange. The second part $\propto\!K$ is the Ising-like `Kitaev'-term which is the signature of the entangled, spin-orbital wave function~\cite{Jackeli:2009p2016,Chaloupka:2013ju,Chaloupka:2010p2461}. 

In the following, we parametrize 
\be
J\!=\!\cos\psi, \quad
K\!=\!\sin\psi~.
\ee
We shall also use a primed coordinate frame $\{\vec{x}',\vec{y}',\vec{z}'\}$ to describe the geometry in real space, with $\vec{x}'$ and $\vec{y}'$ defining the plane of the lattice. The spin-orbit coupling locks this frame to the frame $\{\vec{x},\vec{y},\vec{z}\}$ used for the spin space in (\ref{eq:ham}), in such a way that each bond direction $\bs{\epsilon}$ is perpendicular to the corresponding $\gamma_{\bs{\epsilon}}$-axis, and the plane of the lattice is one of the four $\{111\}$ planes in spin space. The $(111)$ choice corresponds to 
\be
\vec{a}\!=\!a\frac{\vec{z}-\vec{x}}{\sqrt{2}},\quad 
\vec{b}\!=\!a\frac{\vec{x}-\vec{y}}{\sqrt{2}},\quad
\vec{c}\!=\!a\frac{\vec{y}-\vec{z}}{\sqrt{2}},
\ee 
where $a$ is the lattice constant.

The combined spin-orbit symmetry of the Hamiltonian is $D_{3d}$ ($\bar{3}m$), as that of layered compounds of ABO$_2$ type. With the above choice of the lattice plane, the threefold axis $[111]$ of $D_{3d}$ maps the spin components and lattice directions as 
\be\label{eq:sym}
(x,y,z)\!\mapsto\!(y,z,x),\quad 
(\vec{c},\vec{a},\vec{b})\!\mapsto\!(\vec{a},\vec{b},\vec{c}),
\ee
while the $C_2$ axes and the reflection planes of $D_{3d}$ are, respectively, parallel and perpendicular to the lattice bonds.

In addition to the $D_{3d}$ symmetry, the model has also a $D_{2h}$ symmetry in spin space alone, where the role of the inversion generator is played by time reversal, and the three twofold axes point along the cubic axes. These twofold axes map $[111]$ to the remaining three $\langle111\rangle$ axes, meaning that the model has essentially four threefold axes and not one.  This is a key aspect for the correct enumeration of all inequivalent types of $Z_2$ vortices, as we explain in Sec.~\ref{sec:types}.
 
Finally, the model admits a duality transformation, denoted by $H_{zyx}$ in the following, similar to the well-known duality of spin-orbital honeycomb lattice models~\cite{Khaliullin05,chaloupka15}. This is a gauge-like transformation that maps the spins of the four-sublattices of Fig.\,\ref{fig:model}\,(a) to rotated spins $\tilde{\vec{S}}$ as: 
\bea
\begin{array}{c}
\vec{S}_1\!=\!\tilde{\vec{S}}_1, \quad
\vec{S}_2\!=\!(-\tilde{S}_2^x,-\tilde{S}_2^y,\tilde{S}_2^z),\\
\vec{S}_3\!=\!(-\tilde{S}_3^x,\tilde{S}_3^y,-\tilde{S}_3^z), \quad
\vec{S}_4\!=\!(\tilde{S}_4^x,-\tilde{S}_4^y,-\tilde{S}_4^z),\end{array}
\eea
which amounts to a product of $\pi$-rotations around $\vec{z}$, $\vec{y}$ and $\vec{x}$ for the sublattices $2$, $3$ and $4$, respectively. This preserves the form of the Hamiltonian (\ref{eq:ham}), but changes $J\!\mapsto\!J'\!=\!-J$ and $K\!\mapsto\!K'\!=\!2J\!+\!K$, mapping different regions of the phase diagram onto each other, as discussed below. Of particular interest are the two special points with $K'\!=\!0$ (or $\tan\psi\!=\!-2$) where the transformed Hamiltonian is SO(3) symmetric: the $\widetilde{\text{HAF}}$ point $\psi\!=\!\pi\!-\!\arctan{2}$, where $J'\!>\!0$, and the \fmtilde counter point $\psi\!=\!-\!\arctan{2}$, where $J'\!<\!0$.

\begin{figure}[!t]
\includegraphics[width=0.99\columnwidth,trim= 0 0 0 0,clip]{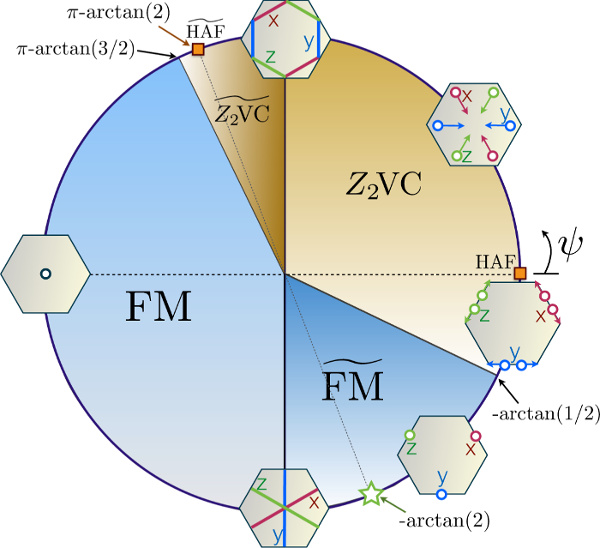}
\caption{The classical $T\!=\!0$ phase diagram of the model (\ref{eq:ham}), parametrized by the angle $\psi$ ($J\!=\!\cos\psi$, $K\!=\!\sin\psi$). There are four extended phases [FM, $\widetilde{\text{FM}}$, $Z_2$VC, and $\widetilde{Z_2\text{VC}}$] plus four isolated phases: the HAF at $\psi\!=\!0$, its dual $\widetilde{\text{HAF}}$ at $\psi\!=\!\pi\!-\!\arctan{2}$, and the two highly degenerate Kitaev points at $\psi\!=\!\pm\pi/2$. The inset hexagons show the positions of the minima of $\lambda_\gamma(\vec{k})$ (colored markers or line segments at $\psi\!=\!\pm\pi/2$) in the first BZ, while arrows indicate how they shift when going around the circle in a counter-clockwise direction. The impact of thermal and quantum fluctuations is discussed in the text.\label{fig:phases}} 
\end{figure}

\section{Global phase diagram}\label{sec:gpd}
The classical $T\!=\!0$ phase diagram of the model is shown in Fig.~\ref{fig:phases} in the whole parameter region $\psi\in[0,2\pi)$. There are four extended phases, FM, $\widetilde{\text{FM}}$, $Z_2$VC, and $\widetilde{Z_2\text{VC}}$, plus the two isolated Kitaev points ($\psi\!=\!\pm\pi/2$), and the two isolated AF Heisenberg points, HAF ($\psi\!=\!0$) and $\widetilde{\text{HAF}}$ ($\pi\!-\!\arctan{2}$). Under $H_{zyx}$, \fmtilde is the dual phase of FM (with $\psi\!=\!-\arctan{2}$ mapping to $\psi\!=\!\pi$), $\widetilde{Z_2\text{VC}}$ is the dual of $Z_2$VC, $\widetilde{\text{HAF}}$ is the dual of HAF, while the two Kitaev points are self-dual.

The four different regions of Fig.~\ref{fig:phases} coincide with the qualitatively different regimes extracted from the LT minimization method~\cite{LT,*Bertaut,*Litvin,*Kaplan}, so let us first discuss this method. Here one replaces the strong spin-length constraints ($\vec{S}_i^2\!=\!1, \forall i$)  of the problem with a single, much weaker constraint, $\sum_i \vec{S}_i^2\!=\!N$, where $N$ is the number of spin sites. The associated linear problem amounts to a straightforward minimization in momentum space with a single Lagrange multiplier. The resulting LT solutions correspond to exact classical minima if they also happen to satisfy the strong constraints, which is not always true in our model as we discuss below.

With $\vec{S}_{\vec{k}}\!=\!\sum_{\vec{r}} e^{i\vec{k}\cdot\vec{r}}\vec{S}_{\vec{r}}$, the total energy can be expressed in terms of a 3$\times$3 coupling matrix $\bs{\Lambda}(\vec{k})$ as
\be\label{eq:LT}
\mc{H}/N=\sum_{\vec{k}}\vec{S}_{\vec{k}}\cdot \bs{\Lambda}(\vec{k})\cdot\vec{S}_{-\vec{k}}~.
\ee
Here $\bs{\Lambda}(\vec{k})$ is diagonal in the basis $\{\vec{x},\vec{y},\vec{z}\}$, with eigenvalues 
\be
\lambda_{\gamma}(\vec{k}) \!=\! K\cos(\vec{k} \cdot \bs{\epsilon}_{\gamma}) \!+\! J\! \sum_{\bs{\epsilon}'\!=\!\bs{a},\bs{b},\bs{c}}
\cos(\vec{k} \cdot \bs{\epsilon}'),
~\gamma\!=\!x,y,z \;. \label{eq:lambda}
\ee
where $\bs{\epsilon}_{x}\!=\!\vec{c}$, $\bs{\epsilon}_{y}\!=\!\vec{a}$ and $\bs{\epsilon}_{z}\!=\!\vec{b}$. 
Minimizing these coupling functions over the first Brillouin zone (BZ) of the model gives the four different regions of Fig.~\ref{fig:phases}. The eigenmodes corresponding to the minima $\lambda_{\text{min}}$ of the three coupling functions (which are degenerate due to the threefold symmetry) satisfy (or can be combined so that they satisfy) the spin length constraint at each site inside the regions FM and $\widetilde{\text{FM}}$, and the same is true for the isolated Kitaev points $\psi\!=\!\pm\pi/2$, as well as for HAF and $\widetilde{\text{HAF}}$. In the remaining parts of the phases $Z_2$VC and $\widetilde{Z_2\text{VC}}$, on the other hand, the minima correspond to three pairs of incommensurate wavevectors $\pm\vec{Q}^{(\gamma)}$ (see inset hexagons of Fig.~\ref{fig:phases}), one for each spin component $\gamma$, and the LT method fails to deliver a state that satisfies the length constraint at each site. In this case, the momenta $\pm\vec{Q}^{(\gamma)}$ are in rough, qualitative agreement with the first harmonics of the actual spin structure factor, while $\lambda_{\text{min}}$ serves as a low bound for the ground state energy.

Let us discuss the different regions in more detail, before we move to the $Z_2$VC phase, which is the main subject of our study.

\begin{figure*}[!t]
\includegraphics[width=0.9\columnwidth,angle=90,trim={3cm 3cm 3cm 3cm}]{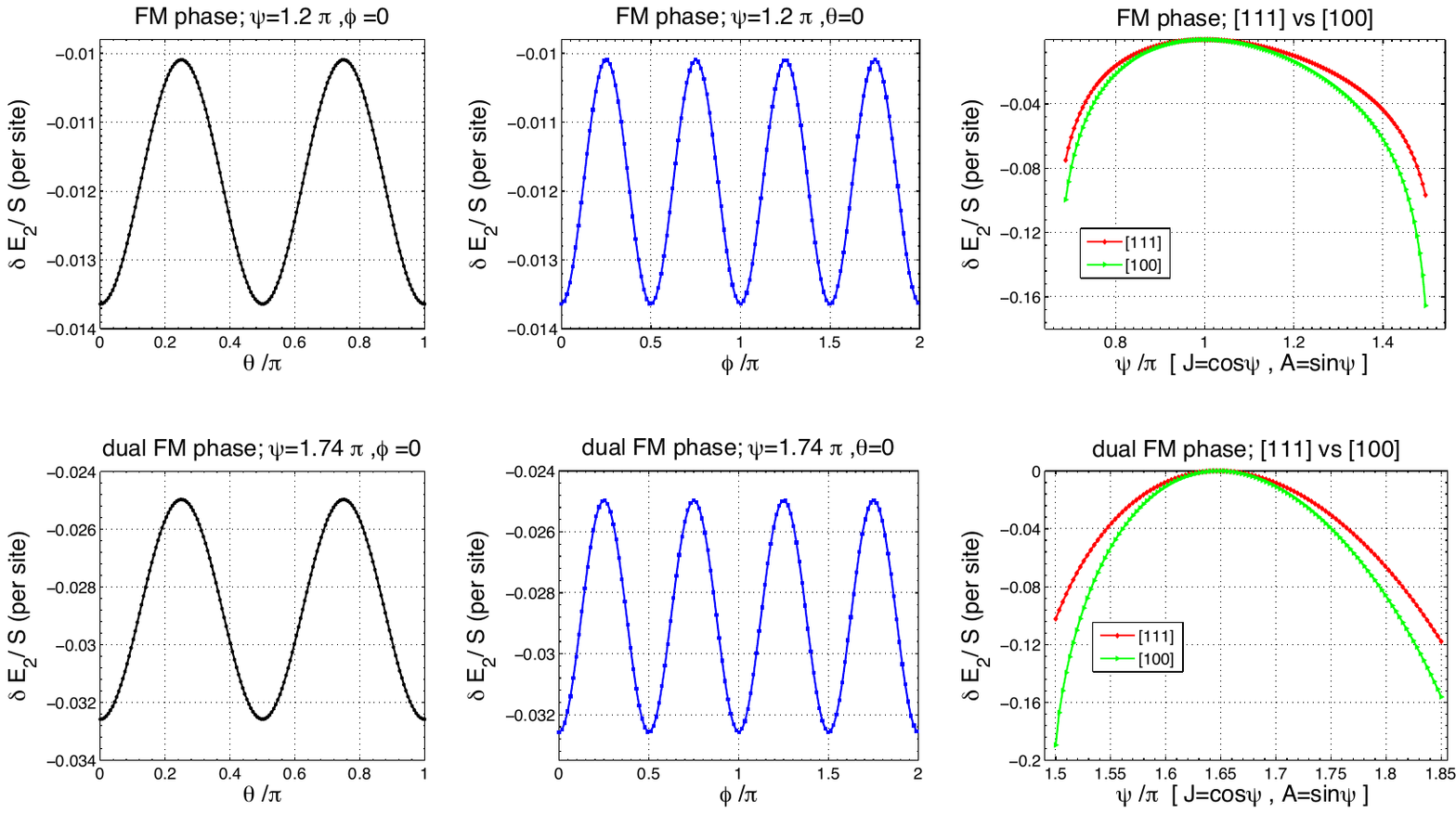}
\caption{Quantum order-by-disorder inside the FM and the \fmtilde regions of the phase diagram of Fig.~\ref{fig:phases}. Here we show the quantum energy correction $\delta E^{(2)}$ (divided by the spin $S$, per site) from non-interacting spin wave fluctuations around different ground states parametrized by the polar and azimuthal angles, $\theta$ and $\phi$, of the FM order parameter. The last column shows a comparison of the energy corrections for ordering along the [111] and along the cubic [100] axis as a function of the coupling parameter $\psi$ of the model.\label{fig:OBD}} 
\end{figure*}

\subsection{FM and \fmtilde phases}\label{sec:FM}
Inside the FM region, the minima of $\lambda_\gamma(\vec{k})$ reside at the $\bs{\Gamma}$ point ($\vec{k}\!=\!0$) of the first BZ. The corresponding solutions are the fully polarized states along the three cubic axes $\vec{x}$, $\vec{y}$, and $\vec{z}$. However, since the three $\lambda_{\text{min}}$ are the same, any global direction in spin space gives the same energy.  The resulting SO(2) manifold of states reads
\be\label{eq:FM111}
\text{FM}: \quad
\vec{S}_{\vec{R}}= \left( f_x,~f_y,~ f_z\right),
\ee
where $f_x^2+f_y^2+f_z^2\!=\!1$. This degeneracy is accidental (except for the SO(3) point $\psi\!=\!\pi$), and should therefore be lifted by fluctuations, see below.

The physics is the same in the \fmtilde region, by virtue of the duality transformation, however the structure of the ground state manifold is much richer when drawn in the unrotated frame. In particular, these states are generally non-coplanar, AF, and have unit cells containing up to four spins, with spin patterns depending on the direction $\tilde{\vec{S}}$ of the spins in the rotated frame. In this region, the minima reside at the three $\vec{M}$-points of the BZ [$\vec{M}^{(x)}\!=\!\frac{1}{a}(\pi,\!\frac{\pi}{\sqrt{3}})$, $\vec{M}^{(y)}\!=\frac{1}{a}\!(0,\!\frac{2\pi}{\sqrt{3}})$, $\vec{M}^{(z)}\!=\!\frac{1}{a}(-\pi,\!\frac{\pi}{\sqrt{3}})$], while the  accidental (except for $\psi\!=\!-\arctan{2}$) SO(3) degeneracy present in the rotated frame, results by combining these special points, which can be done without compromising the spin length constraints~\cite{Villain1977,Fouet}. More explicitly, the SO(2) manifold of degenerate states now reads 
\be\label{eq:tildeFM111}
\widetilde{\text{FM}}: \quad
\vec{S}_{\vec{R}}= \left( f_x~e^{i \vec{M}^{(x)}\cdot\vec{R}},~
f_y~e^{i \vec{M}^{(y)}\cdot\vec{R}},~ f_z~e^{i \vec{M}^{(z)}\cdot\vec{R}}\right),
\ee
where again $f_x^2+f_y^2+f_z^2\!=\!1$. 

Of particular interest are the states in (\ref{eq:tildeFM111}) with $|f_x|\!=\!|f_y|\!=\!|f_z|\!=\!\frac{1}{\sqrt{3}}$, where all spins point along the $\langle111\rangle$-axes. As we discuss in Sec.~\ref{sec:Z2}, this particular states can be thought of as carrying the smallest $Z_2$ vortices for $K\!<\!0$, and are thus selected as soon as we cross the boundary from $\widetilde{\text{FM}}$ to $Z_2$VC. The analogous role of the states in (\ref{eq:FM111}) with $|f_x|\!=\!|f_y|\!=\!|f_z|\!=\!\frac{1}{\sqrt{3}}$, at the boundary between FM and $\widetilde{Z_2\text{VC}}$, follows by duality. These states have the special property that each elementary triangle carries a very large chirality $|\bs{\kappa}|\!=\!8/9$ [see definition (\ref{eq:kappa}) below], and in this sense are the closest to the
chiral 120$^\circ$ state of the HAF point, where $|\bs{\kappa}|\!=\!1$. Note that the coupling of itinerant charge carriers to such a spin pattern gives rise to  topologically non-trivial Chern bands.~\cite{Martin08}

\begin{figure}[!b]
\includegraphics[width=0.35\textwidth,trim=0 0 0 0, clip]{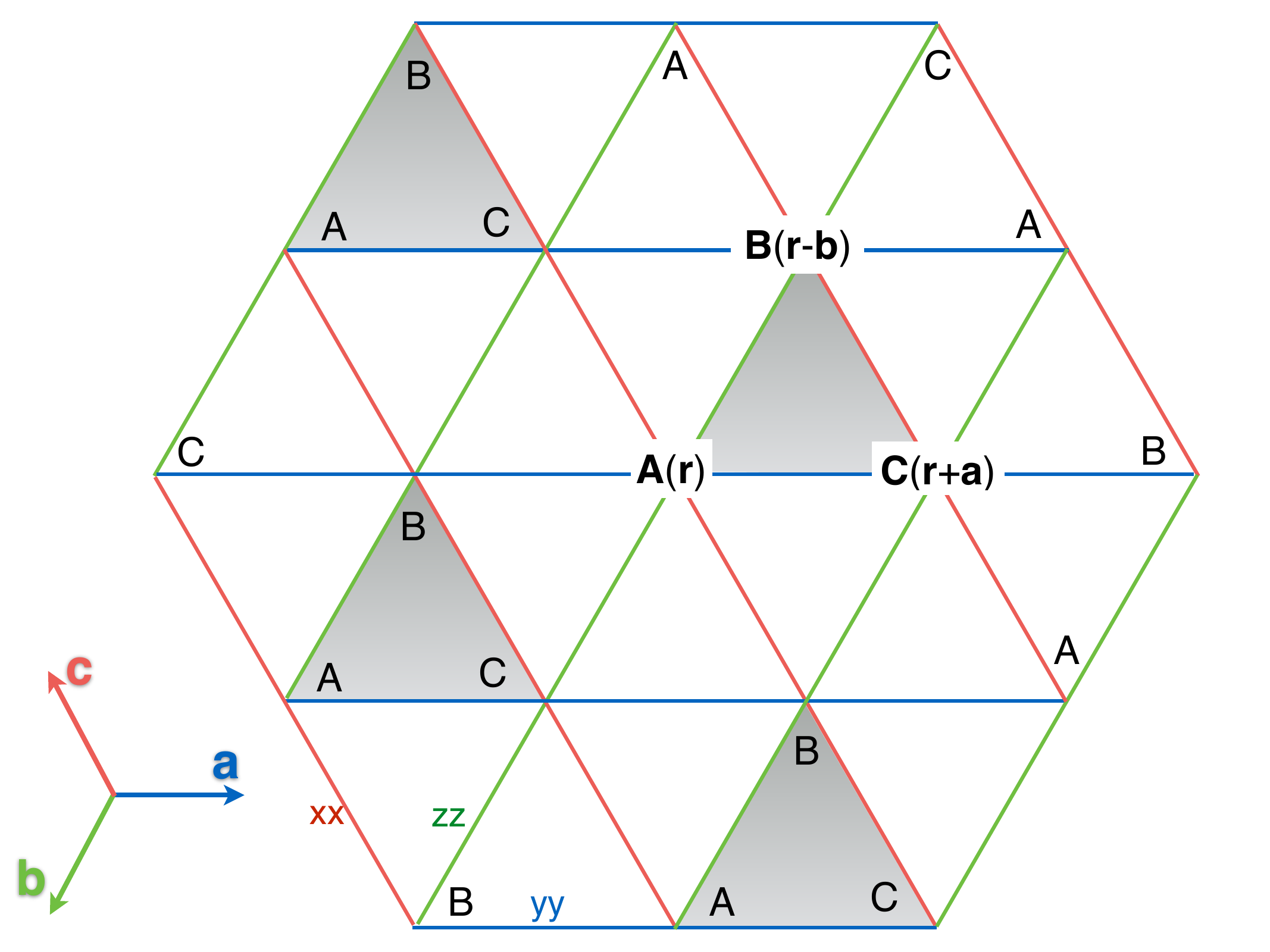}
\caption{The trimerized lattice structure of the 120$^\circ$ state, with the letters A, B and C denoting the three spin sublattices. The corresponding vector fields $\vec{A}(\vec{r})$, $\vec{B}(\vec{r}-\vec{b})$ and $\vec{C}(\vec{r}+\vec{a})$ parametrize the SO(3) order parameter in the long-wavelength expansion of Sec.~\ref{sec:LDT}.}\label{fig:TrimLattice} 
\end{figure}

\subsubsection{Effect of thermal and quantum fluctuations}\label{sec:FMfluctuations}
Except for the special SO(3) points $\psi\!=\!\pi$ and $\psi\!=\!-\!\arctan{2}$, the degeneracy associated with the SO(2) manifold of ground states inside the regions FM and $\widetilde{\text{FM}}$ is accidental and therefore should be lifted by thermal or quantum fluctuations. As shown in Fig.~\ref{fig:OBD}, the zero-point energy from harmonic spin waves selects the cubic axes in both the FM and \fmtilde regions, similarly to what happens in several anisotropic spin-orbital, compass-like models~\cite{Khaliullin01,Mishra:2004p2012,Dorier2005,Nussinov2015}.

As mentioned above and discussed in more detail below, a $\widetilde{\text{FM}}$ state pointing along the $\langle111\rangle$-axes is the same as the $Z_2$VC state with smallest periodicity, allowing a natural connection of the two phases at $\psi\!=\!-\arctan{1/2}$. As quantum fluctuations suppress this orientation on the $\widetilde{\text{FM}}$ side of the phase boundary and favor orientation along cubic axes (and hence a collinear two-site unit cell) they might likewise supress the $\widetilde{Z_2\text{VC}}$ state in the vicinity of $\psi\!=\!-\arctan{1/2}$. As usual, analogous arguments hold for the FM-$\widetilde{Z_2\text{VC}}$ boundary.

\subsection{HAF and $\widetilde{\text{HAF}}$ points}
At the HAF point, $\psi\!=\!0$, the minima of $\lambda_\gamma(\vec{k})$ reside at the corners of the BZ, $\pm\vec{X}\!=\!\pm\frac{1}{a}(\frac{2\pi}{3},\!\frac{2\pi}{\sqrt{3}})$, see Fig.~\ref{fig:model}(b). In real space, the wavevectors can be combined to give the well-known three-sublattice, coplanar $120^\circ$ state of Fig.~\ref{fig:model}(c)~\cite{PhysRev.87.290}, which is known to be stable against quantum fluctuations at zero temperature~\cite{PhysRevLett.71.1629}. Again, the physics is the same at the $\widetilde{\text{HAF}}$ point, $\psi\!=\!\pi-\arctan{2}$, by virtue of the duality $H_{zyx}$, but the structure of the GS manifold is much richer when drawn in the unrotated frame. Here, the three sublattices of the $120^\circ$ state together with the four-sublattice structure of the duality transformation give a twelve-site unit cell with generally non-coplanar spins in the unrotated frame~\footnote{Spins $\vec{S}$ are only coplanar if the plane formed by the rotated spins $\tilde{\vec{S}}$ is orthogonal to a NN bond}. This state has also been discussed in Refs.~[\onlinecite{Khaliullin05,Reuther:2012vf,Kargarian12}]. 

The physics becomes much more interesting as soon as we depart from these points, whereby the minima of $\lambda_\gamma(\vec{k})$ shift to incommensurate wave-vectors, see Sec.~\ref{sec:Z2}.

\begin{figure}[!t]
\subfigure[]{\includegraphics[width=0.45\columnwidth, trim= 25 30 40 40, clip]{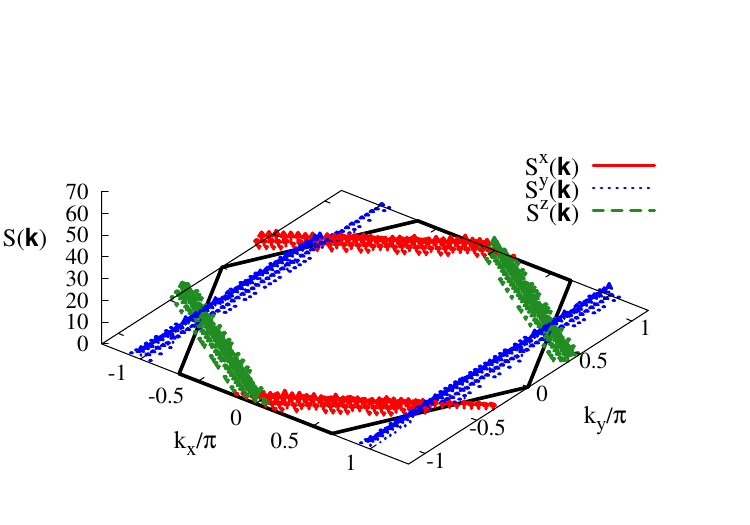}\label{fig:Sk_H0_K1_b8}}\subfigure[]{\includegraphics[width=0.45\columnwidth, trim= 25 30 40 40, clip]{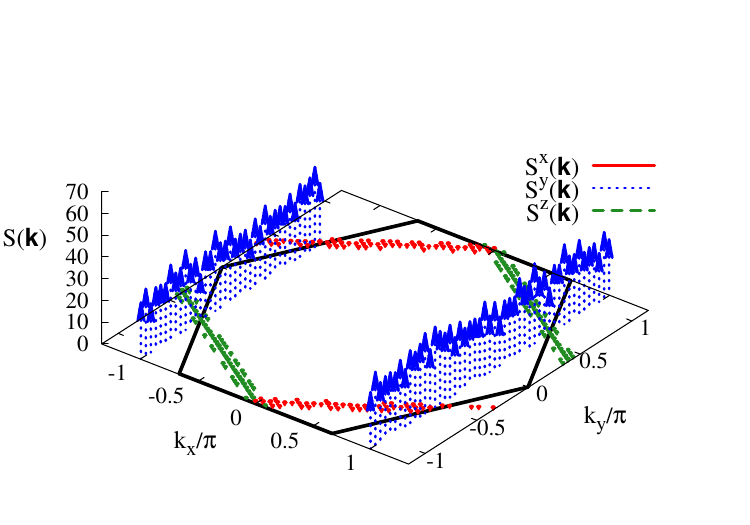}\label{fig:Sk_H0_K1_b10}}\\
\subfigure[]{\includegraphics[width=0.45\columnwidth, trim= 25 30 40 50, clip]{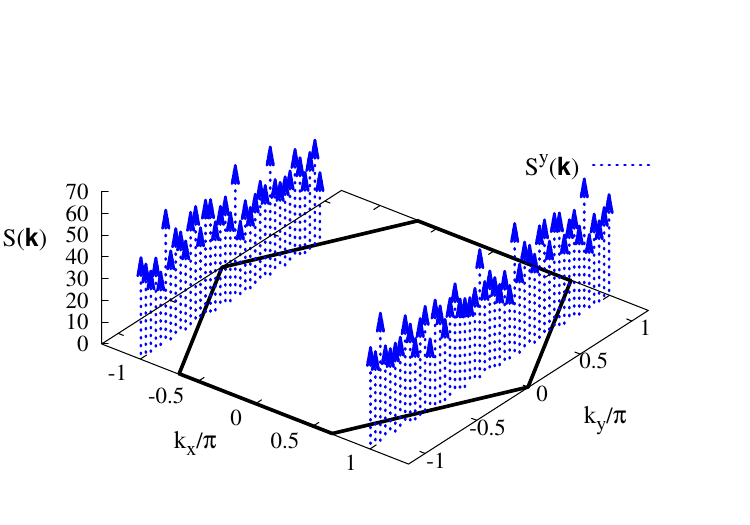}\label{fig:Sk_H0_K1_b100}}\subfigure[]{\includegraphics[width=0.45\columnwidth, trim= 25 30 40 50, clip]{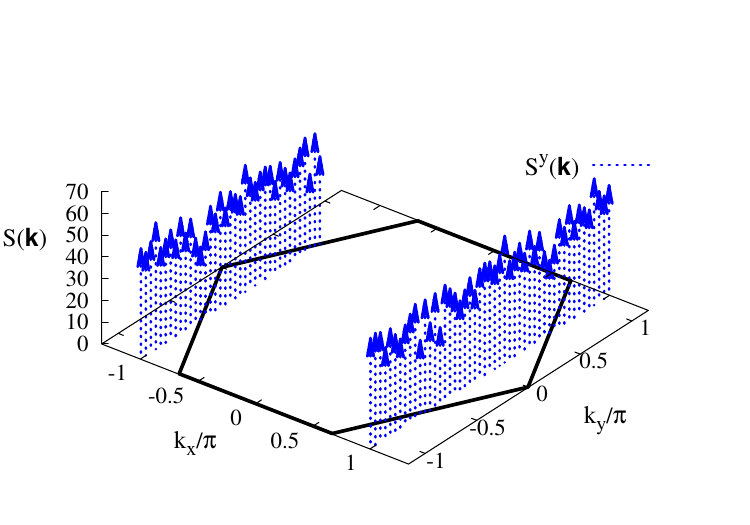}\label{fig:Sk_H0_K1_b1000}}
\caption{Nematic ordering at the AF Kitaev point $\psi\!=\!\pi/2$, as found by classical Monte Carlo simulations of a 48$\times$48 cluster. The data show the three spin structure factors $\mc{S}^{\gamma}({\bf k})\!\equiv\!|S_{\bf k}^\gamma|^2$ at inverse temperature (a) $\beta=8$, where signals in all three components are visible, (b) $\beta\!=\!10$, (c) $\beta\!=\!100$, and (d) $\beta\!=\!1000$, where decoupled chains along one direction dominate.  Momenta ${\bf k}$ are rescaled by a factor of $4a/3$.\label{fig:Sk_H0_K1}}
\label{fig:nematic}
\end{figure}

\subsection{Kitaev points}\label{sec:kitaev}
The physics at the pure Kitaev points, $\psi\!=\!\pm\frac{\pi}{2}$, is very similar to the physics of anisotropic, compass-like models~\cite{Doucot2005,Mishra:2004p2012,wenzel:064402,Dorier2005,Nussinov2015}. First, the minima of $\lambda_{\gamma}(\vec{k})$ form whole lines in the BZ (see Fig.~\ref{fig:phases}), suggesting a ground state manifold with a sub-extensive degeneracy. Indeed, it is easy to show that at $\psi\!=\!\frac{\pi}{2}$ ($-\frac{\pi}{2}$), the lower energy bound $E_0\!=\!-|K|$ is saturated by forming AF (FM) Ising chains along one of the three lattice directions, with spins pointing along the corresponding cubic axis, ${\bf x}$, ${\bf y}$, or ${\bf z}$. Flipping one spin component of all spins along a single chain with the corresponding Ising coupling, does not change the energy, because neighboring chains couple only via the remaining two components. These {\it sliding} operations~\cite{Batista05,Nussinov2005,Nussinov2006,Nussinov2015} lead to $3\times 2^L$ states, where $L$ is the linear system size, hence the subextensive structure of the ground state manifold.

More specifically, for $\psi\!=\!\frac{\pi}{2}$ (the situation at $\psi\!=\!-\frac{\pi}{2}$ is similar) the three families of states can be written as  
\bea
\vec{S}^{(x)}_{\vec{r}} \!=\! x_{n-m} (-1)^m \vec{x}, ~~
\vec{S}^{(y)}_{\vec{r}} \!=\! y_m (-1)^n \vec{y},~~
\vec{S}^{(z)}_{\vec{r}} \!=\! z_n (-1)^m \vec{z},~~
\eea
where the lattice points $\vec{r}\!=\!n\vec{a}+m\vec{b}$ and the sets $\{x_m\}$, $\{y_m\}$ and $\{z_m\}$ are random choices of $\pm1$. We emphasize that the superscripts $(x)$, $(y)$, and $(z)$ do not denote the components of the spins, but index three different families of states. Each state of a given family $(\gamma)$ can be formed by combining the different modes along the respective lines of minima of $\lambda_\gamma(\vec{k})$. 

The above states are actually connected by other, continuous valleys of states which are generated by combining all three lines of minima, leading to a SO(2) manifold of ground states. These states are of the form 
\be\label{eq:Kitaev111}
\text{AF Kitaev:}\quad 
\vec{S}_{\vec{r}} \!=\! \Big(
f_x  x_{n-m} (-1)^m,
f_y y_m (-1)^n, 
f_z  z_n (-1)^m \Big),
\ee
where $f_x^2+f_y^2+f_z^2\!=\!1$. Clearly, this degenerate manifold contains not only collinear but also coplanar and non-coplanar states. 

Again, of particular interest are the states with
$|f_x|\!=\!|f_y|\!=\!|f_z|\!=\!\frac{1}{\sqrt{3}}$, with all spins
pointing along the $\langle111\rangle$ axes. These arise by combining
the centers of each line of minima,
$Q^{(x)}\!=\!\frac{1}{a}(\frac{\pi}{2},\!-\frac{\sqrt{3}\pi}{2})$,
$Q^{(y)}\!=\!\frac{1}{a}(-\pi,\!0)$, and
$Q^{(z)}\!=\!\frac{1}{a}(\frac{\pi}{2},\!\frac{\sqrt{3}\pi}{2})$,
which are commensurate with the lattice. As we discuss in
Sec.~\ref{sec:Z2}, these states carry the smallest $Z_2$ vortices for
$K\!>\!0$. As for $K\!<\!0$, however, fluctuations have to be taken into account
when discussing states chosen from a degenerate manifold.

\subsubsection{Effect of thermal fluctuations}\label{sec:KitaevThermal}
Classically the degeneracy associated to the above SO(2) manifold of states is accidental and as such it can be lifted by thermal fluctuations (quantum fluctuations will be discussed separately below). Our Monte-Carlo (MC) data of Fig.~\ref{fig:Sk_H0_K1} show clearly that this accidental degeneracy is lifted by thermal fluctuations, leading to a finite-$T$ nematic phase where spins select spontaneously one of the three directions of the lattice and fluctuate in the corresponding cubic axis direction in spin space. At finite temperatures $T$, this phase cannot have long-range magnetic order due to the generalized Elitzur's theorem of Batista and Nussinov~\cite{Batista05} which, for the present situation, asserts that the sliding symmetries cannot break spontaneously at any finite $T$~\footnote{The sliding symmetries can break spontaneously only at $T\!=\!0$ and in all possible ways, which is reflected in the divergence of the spin structure factor along lines in momentum space.}. Such partial nematic order into decoupled chains is known to emerge in the low-energy limit of more complex models~\cite{PhysRevB.84.024408,*PhysRevLett.107.076405} and is again analogous to the situation in the square-lattice `compass' model~\cite{Mishra:2004p2012,wenzel:064402}. 

Importantly, the fact that thermal fluctuations select the cubic axes and not the $\langle111\rangle$ axes suggests that, at finite $T$, the nematic phase at the AF Kitaev point can survive in a finite window around $\psi\!=\!\pi/2$, provided that thermal fluctuations inside the neighboring $Z_2$VC and $\widetilde{Z_2\text{VC}}$ phases are weak enough. This is indeed what we find in our MC simulations for $\tan\psi\!=\!4$, $10$ and $20$ (not shown), which reveal that the partial order into decoupled AF chains found at $\psi\!=\!\tfrac{\pi}{2}$  (Fig.~\ref{fig:nematic}) persists over wide intermediate temperature ranges. 

The situation around the FM Kitaev point is different. Unlike the AF Kitaev point where 2D order is strongly frustrated, here even a weak $J\!\neq\!0$ couples the FM chains with an energy $\propto J L$. So the finite-$T$ nematic physics is only present at $\psi\!=\!-\pi/2$, and as soon as we depart from this point the system enters the FM or \fmtilde phase. 

\subsubsection{Effect of quantum fluctuations}\label{sec:KitaevQuantum}
The effect of quantum fluctuations is qualitatively different from the thermal case, but also from the corresponding quantum case in the square lattice compass model. The reason is that here the sliding symmetries do not exist for quantum spins, because flipping one component of the spin requires the use of the time-reversal operation, which cannot be made to act locally, on a single chain. In the square lattice, this obstacle is avoided because the Hamiltonian contains only two types of Ising couplings, and the sliding operations can be effected by $\pi$-rotations. This fundamentally different symmetry structure of the classical and quantum Hamiltonians on the triangular lattice has very striking ramifications for quantum spins: The system can develop long-range magnetic order even at finite $T$. Essentially, this means that neighboring chains couple to each other by virtual quantum-mechanical processes. This has been nicely shown analytically by Jackeli and Avella~\cite{jackeli15} (a similar analysis is carried out in a related honeycomb lattice model~\cite{ioannisK1K2}), and numerically by Becker {\it et al}~\cite{trebst14}. 

Similar to the thermal case discussed above, the energy gain associated with this quantum order by disorder mechanism can stabilize the magnetic LRO phase in a finite window around the AF Kitaev point, provided the quantum fluctuations of the neighboring phases $Z_2$VC and $\widetilde{Z_2\text{VC}}$ are weak enough. This scenario is confirmed by the numerical results of Becker {\it et al}~\cite{trebst14}.

\begin{figure}[!t]
\subfigure[~Spin pattern]{\includegraphics[width=0.52\columnwidth, trim = 65 70 50 60, clip]{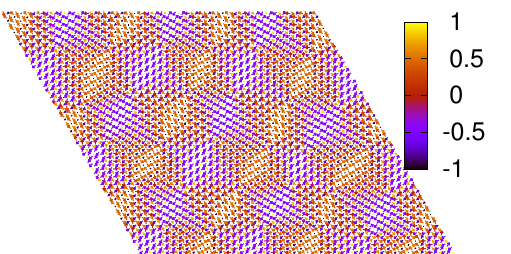}\label{fig:spins_H1_Km025}}
\subfigure[~Spin structure factors]{\includegraphics[width=0.47\columnwidth, trim= 25 20 0 0, clip]{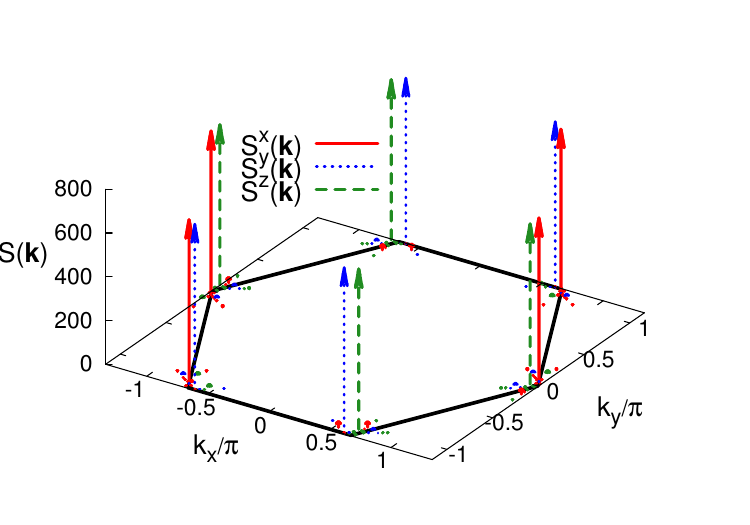}\label{fig:Sk_H1_Km025}}
\\
[-1.3em]
\subfigure[~Local energy]{\includegraphics[width=0.51\columnwidth, trim = 67 50 51 60, clip]{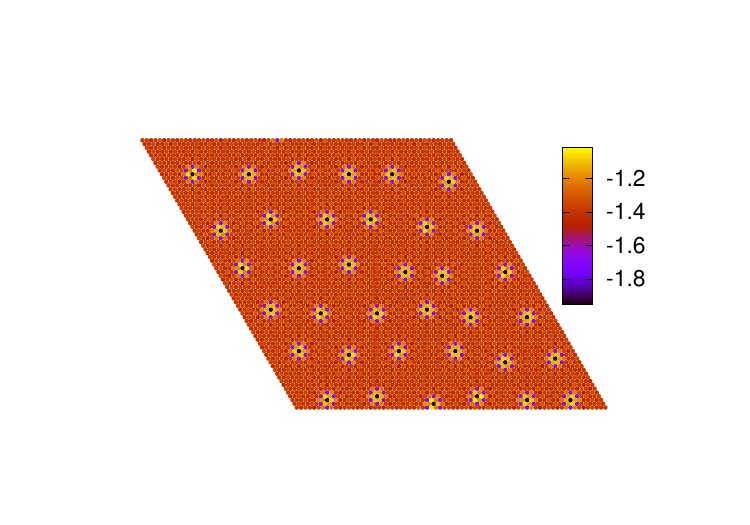}\label{fig:emap_H1_Km025}}
\subfigure[~FT of energy pattern]{\includegraphics[width=0.47\columnwidth, trim= 30 20 30 20, clip]{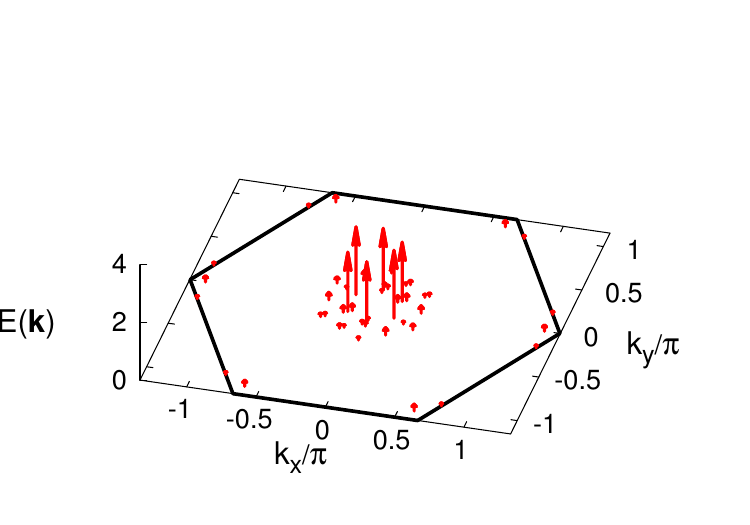}\label{fig:Ek_H1_Km025}}
\caption{(a) Non-coplanar configuration obtained by MC for $K/J\!=\!-0.25$ and $72\times 72$ sites. Shading refers to the out-of plane component from $-1$ (dark) to $1$ (light yellow). 
(b) Corresponding spin structure factors $\mc{S}^\gamma({\bf k})$ in the first BZ. Only peaks with a weight $\geq 1$ are shown.
(c) Local energy $e_{{\bf r}_i} \!=\! J\sum_{\bs{\delta}}{\bf S}_{{\bf r}_i}\!\cdot\!{\bf S}_{{\bf r}_i+{\bs{\delta}}}\!+\!K\sum_{\gamma,{\bs{\delta}}\parallel\bs{\epsilon}_\gamma}\!S^{\gamma}_{{\bf r}_i}\! S^{\gamma}_{{\bf r}_i+{\bs{\delta}}}$, where $\bs{\delta}$ denotes NN bonds.
(d) Fourier transform of $|e_{{\bf r}_i}\!-\!\bar{e}|$, where $\bar{e}$ is the average energy per site. Only peaks with a weight $\geq 0.1$ are shown. 
Momenta ${\bf k}$ are rescaled by a factor of $4a/3$ in (b) and (d).
}
\label{fig:mcmc_spins}
\end{figure}

%-----------------------------------------------------------------------------------
\section{The $Z_2$VC phase}\label{sec:Z2}
We now move to the central topic of our study which is the $Z_2$VC phase. This section is divided into several parts, each one focusing on a different qualitative aspect of this phase. We shall begin by analyzing numerical results from MC simulations (Secs.~\ref{sec:Incom}-\ref{sec:Z2nature}), which were the first to reveal the structure of the $Z_2$ vortices. Building on this knowledge we shall then (Secs.~\ref{sec:domains}-\ref{sec:SmallestVortices}) construct numerically (as described in Sec.~\ref{sec:dv}) and analyze `optimal' $Z_2$VC states in order to get much better ground state energies and much more accurate predictions for the vortex distance as a function of $K/J$. We shall focus entirely on the $Z_2$VC phase, since the physics of the $\widetilde{Z_2\text{VC}}$ phase follows by duality.

%-----------------------------------------------------------------------------------
\begin{figure*}[!t]
\subfigure[]{\includegraphics[width=0.8\columnwidth,trim=0 65 0 50 ,clip]{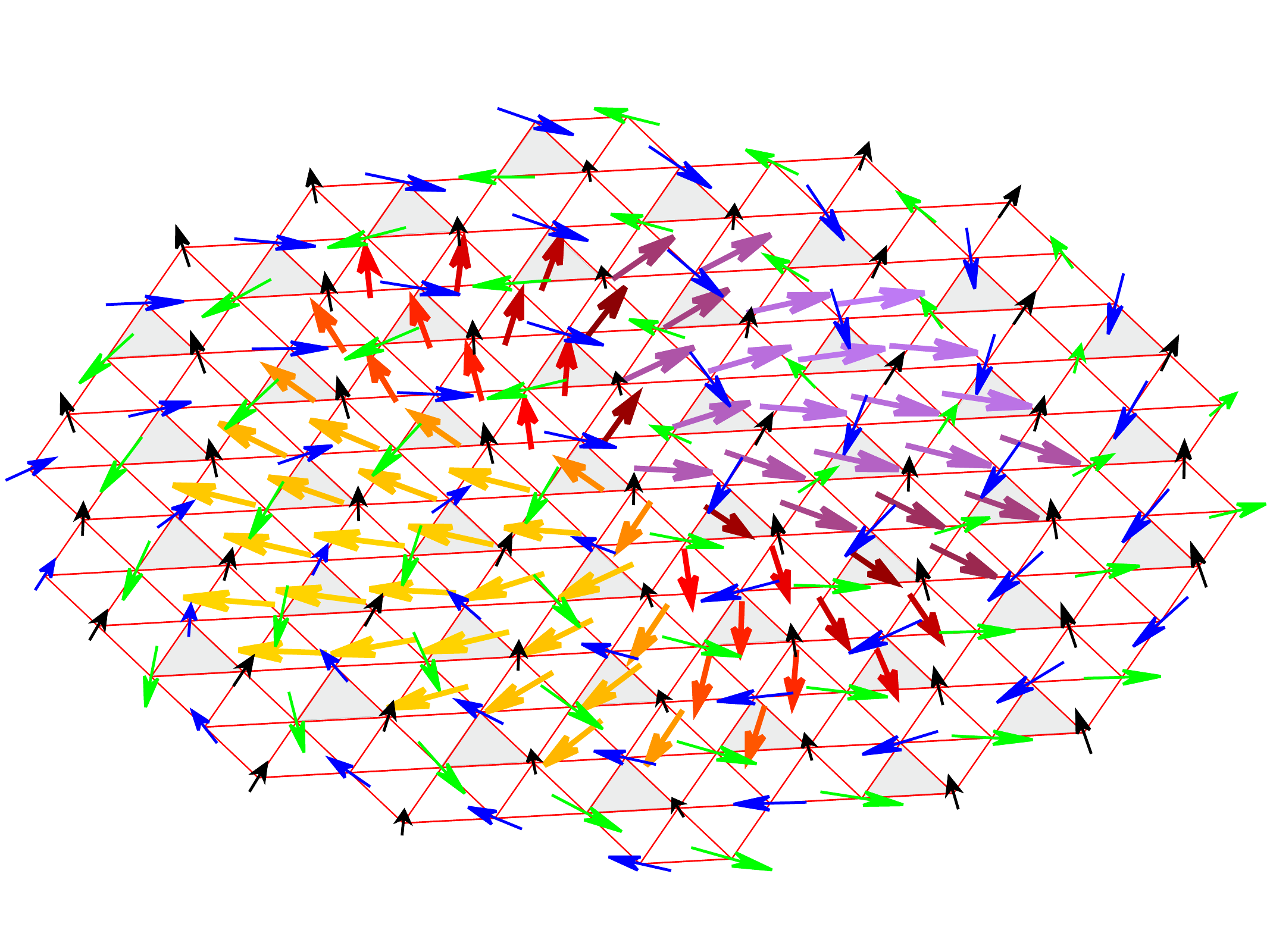}\label{fig:vortex}}
\hspace*{-0.08\columnwidth}
\subfigure[]{\includegraphics[width=0.85\columnwidth, trim = 67 50 50 65, clip]{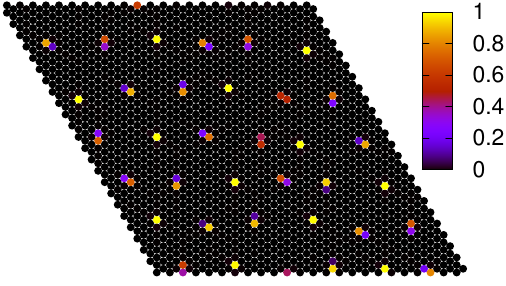}\label{fig:vort_H1_Km025}}\\[-0.5em]  
\caption{The $Z_2$ nature of the cores. (a) A single $Z_2$ vortex on the triangular lattice, obtained in a Monte-Carlo simulation for $K\!=\!-0.25J\!<\! 0$ and 72$\times$72 sites. Black, green and blue arrows on the lattice sites   (residing clockwise around each shaded triangle) indicate spins in the three sublattices of the 120$^\circ$ state. The remaining arrows, that reside on the upward pointing triangles and are shown with a varied color scheme (according to their direction), represent the vector chirality $\bs{\kappa}({\bf r})$,  i.e., they are perpendicular to the plane of the local $120^\circ$ pattern of the plaquette. These vectors form a vortex and almost exactly lie in one plane; In terms of the spins, here the plane formed by two sublattices (blue and green) rotates by $2\pi$ around the third (black), which is roughly constant in this region. For visibility, a global rotation was applied to the spins to make the $\bs{\kappa}$ plane coincide with the  lattice plane, the $\bs{\kappa}$ plane is actually perp. to $[\bar{1}1\bar{1}]$. (b) Vorticity of the vector chirality $\bs{\kappa}({\bf r})$, indicating where the plane of the local $120^\circ$ order rotates by $2\pi$.} 
\label{fig:kappa_vort}
\end{figure*}

\subsection{Incommensurate and non-coplanar nature}\label{sec:Incom}
As mentioned briefly above, as soon as we depart from the AF point $\psi\!=\!0$, the minima of $\lambda_{\gamma}(\vec{k})$ shift away from the corners of the BZ, to incommensurate momenta. The crucial point is that due to the anisotropy in the Kitaev terms, the shift differs for the three spin components, i.e., their couplings (\ref{eq:lambda}) are optimized by three different ordering momenta $\vec{Q}^{(\gamma)}$, as illustrated in Fig.~\ref{fig:model}(b). This is in contrast to incommensurability induced by \emph{spin-isotropic} frustration, e.g. via longer-range Heisenberg couplings~\cite{Okubo:2012cn}, where the same ordering momenta would optimize all three $\lambda_{\gamma}(\vec{k})$. A coplanar spiral with one of the optimal momenta would then be a classical ground state.  Here, however, no $\vec{Q}^{(\gamma)}$ can minimize simultaneously all three (or even two) coupling functions $\lambda_{\gamma}(\vec{k})$ and the ground state is thus not automatically given by a single ordering momentum. 

To find the ground-state ordering, we proceed by classical Monte-Carlo simulations. At $K\!=\!0$, we find the expected $120^\circ$ pattern where the spin structure factor $\mc{S}^{\gamma}({\bf k})\!\equiv\!|S_{\bf k}^\gamma|^2$ is peaked at the corners of the BZ, $\pm\vec{X}$, for all $\gamma$. At small finite $|K|/J$, spins remain locally close to the $120^\circ$ pattern, but this is distorted at larger distances, resulting in an incommensurate non-coplanar configuration, see the MC data for $K/J\!=\!-0.25$ and 72$\times$72 sites shown in Fig.~\ref{fig:spins_H1_Km025}. The non-coplanarity is also revealed by the ``spin-inertia'' tensor $\varmathbb{I}$, whose elements are defined as~\cite{Henley1984}: 
\be
\varmathbb{I}^{\alpha\beta}\!=\!\sum_i S_i^\alpha S_i^\beta/N~.
\ee 
For a collinear (resp. coplanar) pattern, two (resp. one) of the eigenvalues of $\varmathbb{I}$ must vanish, but all three are here equal to  $1/3$ (with small deviations due to numerical fluctuations and finite-size effects). This is the first indication that the state preserves the threefold symmetry of the problem.

The equivalent role of all three spin components and the incommensurate order also reveal themselves in the spin structure factor data shown in Fig.~\ref{fig:Sk_H1_Km025}. The dominant peaks of $\mathcal{S}^\gamma({\bf k})$ indeed each move slightly away from $\pm\vec{X}$, but all three are present, both in expectation values averaged over the MC run and in single snapshots. These peaks roughly track the positions of the minima of $\lambda_\gamma(\vec{k})$, with each spin component modulating at a different wavevector. As for the spin length constraints, these are eventually satisfied by  the presence of higher harmonics, see App.~\ref{app:HarmAmps}.

\subsection{Particle-like modulations}\label{sec:particles}
Fig.~\ref{fig:spins_H1_Km025} gives in addition the first hint for the presence of localized, particle-like modulations. The local energy profiles shown in Fig.~\ref{fig:emap_H1_Km025} clearly reveal the cores of these modulations, which are arranged in an approximate triangular superlattice. The cores can also be seen in the vorticity of the vector chirality and in the FM canting out of the local 120$^\circ$ structure, see below. The superlattice becomes apparent by a Fourier transform of the deviation of the energy from its average value, which is seen in Fig.~\ref{fig:Ek_H1_Km025}. Such superlattices are also found at other values of $K/J$, both positive and negative, with larger $K$ inducing larger deviations from the $120^\circ$ pattern and denser packing of cores.

Plotting the profiles of the Heisenberg and Kitaev energy contributions separately (not shown) reveals that the former is overall positive around the cores while the latter is negative. So the cores are energetically favored by the Kitaev anisotropy.

\begin{figure*}[!t]
\includegraphics[width=0.96\textwidth,angle=0,clip]{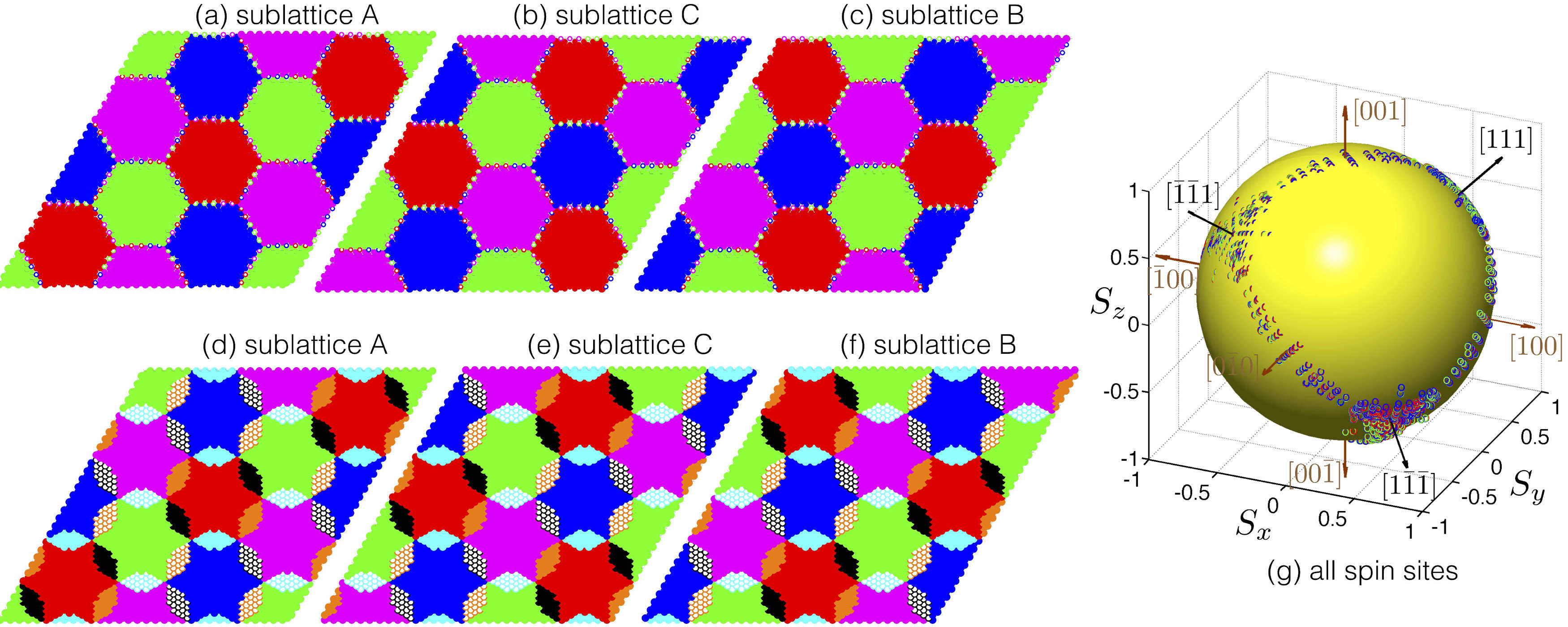}
\caption{Data from the optimal $Z_2$VC for $K/J\!=\!-0.25$ and 96$\times$96 sites (for which $d\!=\!16$, see Table~\ref{tab:Z2data}). (a-c) Largest projection among the ones along the symmetry axes $[111]$ (red), $[1\bar{1}\bar{1}]$ (green), $[\bar{1}1\bar{1}]$ (blue), and $[\bar{1}\bar{1}1]$ (magenta), for each of the three sublattices separately. (d-f) As above, but now we include in the set of projections the ones along the cubic axes $[100]$ (black), $[010]$ (cyan), and $[001]$ (orange). Filled (empty) symbols correspond to positive (negative) values. (g) Projection of all spins in a common SO(2) space.\label{fig:111}} 
\end{figure*}

\subsection{Topological $Z_2$ vortex nature}\label{sec:Z2nature}
We are now going to show that the particle-like modulations correspond to topological $Z_2$ vortices. Such vortices are well known~\cite{Kawamura:1984hk,Kawamura:2010di,Kawamura:2011bq,Tamura12} to be thermally induced above the $120^\circ$ state (or similar AF states~\cite{PhysRevB.77.172413}), since the order parameter has an SO(3) space with the first homotopy group~\cite{Mermin1979} $\pi_1(\text{SO(3)})\!=\!Z_2$. Here, the spin patterns are locally close to the $120^\circ$ state for $|K|\!\ll\!J$, and so the presence of such vortices in the ground state would suggest that the Kitaev anisotropy plays a non-trivial role.

In a $Z_2$ vortex, the plane containing the locally coplanar $120^\circ$ order rotates by $2\pi$, naturally inducing a globally non-coplanar pattern~\cite{Kawamura:1984hk}. The orientation of the  plane is captured by the vector chirality 
\be\label{eq:kappa}
\bs{\kappa}({\bf r}) \!=\! \frac{2}{3\sqrt{3}} ({\bf S}_{{\bf r}}\times{\bf S}_{{\bf r}+{\bf a}}
 +{\bf S}_{{\bf r}+{\bf a}}\times{\bf S}_{{\bf r}-{\bf b}}+{\bf S}_{{\bf r}-{\bf b}}\times{\bf S}_{\bf r}),
\ee
obtained from the three spins around upwards pointing triangles. The prefactor in (\ref{eq:kappa}) sets $|\bs{\kappa}({\bf r})|\!=\!1$ in the $120^\circ$ state. Fig.~\ref{fig:vortex} shows the behavior of the spins in the vicinity of a single core of Fig.~\ref{fig:emap_H1_Km025}. In this region, the spins in one of the three sublattices (black arrows) remain roughly parallel to each other, while the spins in the other two sublattices (blue and green arrows) rotate around the former, in such a way that the vector chirality $\bs{\kappa}({\bf r})$ (represented by an arrow on every upward pointing triangle) completes a $2\pi$ rotation around the core, which in turn proves its topological $Z_2$ nature. The same is true for every core of Fig.~\ref{fig:emap_H1_Km025}, as can be shown by calculating the vorticity of $\bs{\kappa}({\bf r})$ throughout the system, see Fig.~\ref{fig:vort_H1_Km025}.

\subsection{Defected SO(3) nature of the cores}\label{sec:cores}
The cores of the $Z_2$ vortices are defects of the local 120$^\circ$ structure. Indeed, the spins in the immediate surrounding of a core show a finite FM canting. Such a canting is in fact the most natural way to sustain the $Z_2$ vortices in the present lattice model, and can be seen in two different ways. First, the length of the chirality vector $\bs{\kappa}$ becomes smaller than one at the cores. Our results give $|\bs{\kappa}|\simeq0.87$ at the cores (and $|\bs{\kappa}|\simeq1$ away from them), and this amount of reduction remains robust everywhere inside the $Z_2$VC region, see Table~\ref{tab:Z2data} below. 

The second way to see the canting is by looking at the total moment on upward triangles: 
\be
\vec{M}(\vec{r}) = {\bf S}_{{\bf r}}+{\bf S}_{{\bf r}+{\bf a}}+{\bf S}_{{\bf r}-{\bf b}}~.
\ee
Our results confirm that the lengths of these vectors become finite at the cores of the $Z_2$ vortices, with the largest values at the center of the cores being of the order of $|\vec{M}|\!\sim\!1$. This number also remains robust everywhere inside the $Z_2$VC region, see Table~\ref{tab:Z2data} below.

Now, the moment $\vec{M}(\vec{r})$ is not entirely parallel to the chirality vector $\bs{\kappa}(\vec{r})$, meaning that not only the out-of-plane but also the in-plane canting is finite. As we show in Sec.~\ref{sec:LDT}, the presence of a finite in-plane canting can be actually predicted by the form of the long-distance action, which contains a linear derivative cross-coupling term between the in-plane canting and the twisting of the SO(3) order parameter, see Sec.~\ref{sec:FMcanting}.

\subsection{Sublattice FM domain picture: The special role of the $\langle111\rangle$ and the $\langle100\rangle$ axes}\label{sec:domains}
In the vortex shown in Fig.~\ref{fig:vortex} the chirality $\bs{\kappa}(\vec{r})$ lies almost exactly in the plane perpendicular to $[\bar{1}1\bar{1}]$, which is one of the four $\langle111\rangle$ symmetry axes of the model, as we discussed in Sec.~\ref{sec:model}. It turns out that every single $Z_2$ vortex of the ground state is associated with one of these special axes. This is demonstrated in Figs.~\ref{fig:111}~(a-c), which show the maximum among the projections of the spins along $[111]$, $[1\bar{1}\bar{1}]$, $[\bar{1}1\bar{1}]$, and $[\bar{1}\bar{1}1]$, for each of the three sublattices separately. 

These projections reveal that each sublattice of the 120$^\circ$ state (A, B and C, defined as in Fig.~\ref{fig:TrimLattice}) forms a hexagonal superlattice of FM domains with spins pointing roughly (see below) along one of the four $\langle111\rangle$ axes. The key aspect that gives the $Z_2$ vortices is that the A-, B- and C-superlattices are mutually shifted in such a way that the centers of the hexagonal plaquettes, say in A, coincide with vertices in B and C. And since each vertex is the merging point of three hexagonal plaquettes (or domains), it follows that B and C pass sequentially through the other three symmetry axes $[1\bar{1}\bar{1}]$,  $[\bar{1}1\bar{1}]$, and  $[\bar{1}\bar{1}1]$ as we go around the center of the `A-$[111]$' domain. So the plane formed by B and C completes a $2\pi$ rotation around $[111]$, meaning that the core of each hexagonal domain is associated with a $Z_2$ vortex.

Now, the role of the cubic axes comes into light when we examine more closely the rotation of the spins from one symmetry axis to another in the boundary regions between the plaquettes. This is demonstrated in Figs.~\ref{fig:111}~(d-f), that show the maximum among seven projections, the four along $\langle111\rangle$ and the three along the $\langle100\rangle$ axes. In short: i) Spins that reside at the cores of a given sublattice plaquette point exactly along one of the $\langle111\rangle$ axes. ii) Spins that reside on the edges of the plaquettes point exactly along one of the three cubic axes $\langle100\rangle$. iii) Spins away from the cores and the edges follow special paths in SO(2) space, that pass closely to both $\langle100\rangle$ and $\langle111\rangle$ axes. This is demonstrated in Fig.~\ref{fig:111}~(g), which shows the directions of all spins in a common SO(2) space.

\subsection{Solitonic nature of the cores}\label{sec:soliton}
\begin{figure}[!t]
\includegraphics[width=0.4\textwidth,clip]{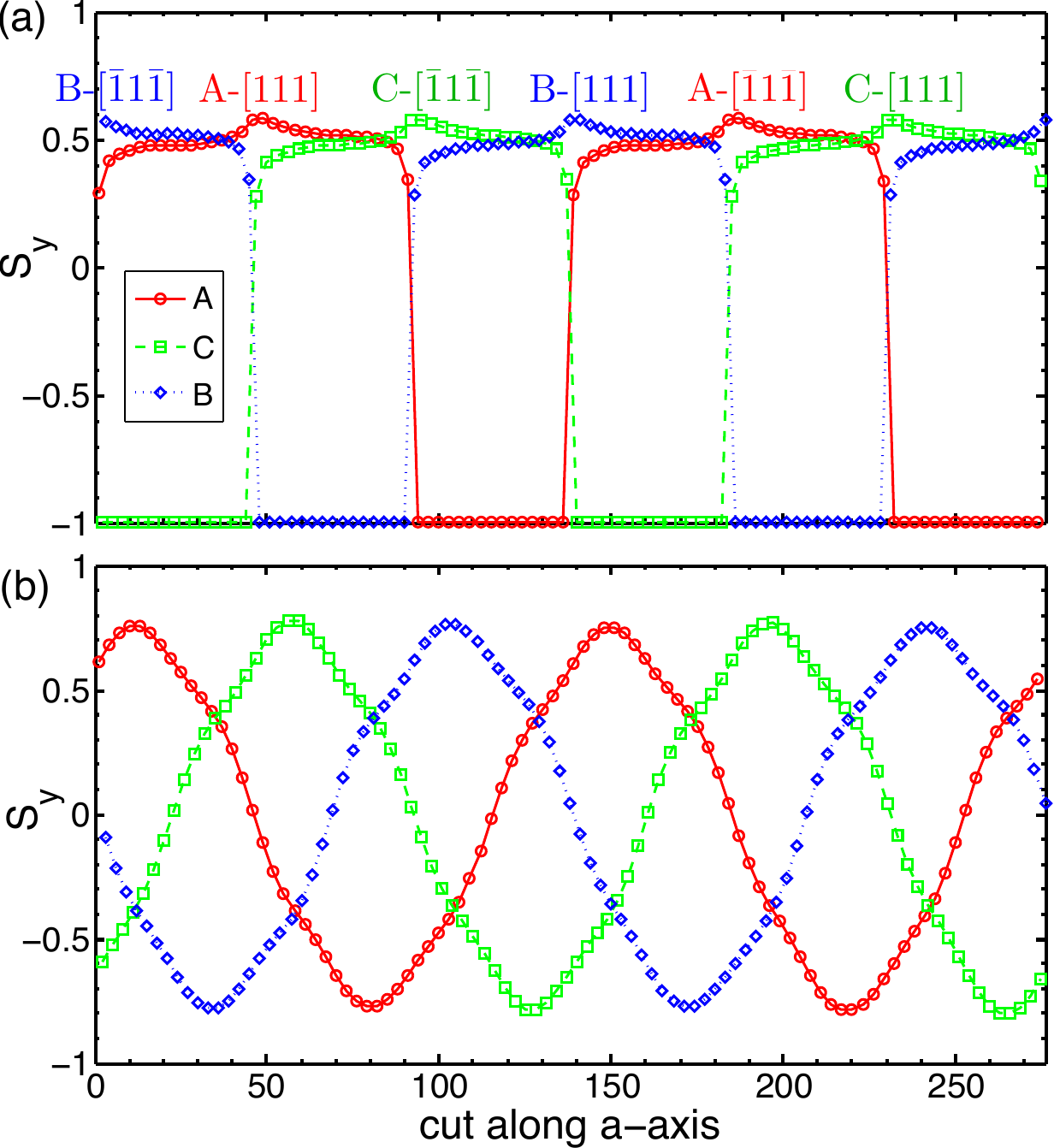}
\caption{Data from the optimal $Z_2$VC for $K/J\!=\!-0.13$ and 276$\times$276 sites (for which $d\!=\!46$, see Table~\ref{tab:Z2data}). Behavior of $S^y$ along the $\vec{a}$ direction, through a cut of the lattice that passes through the cores (a) and half-way between the cores (b). The three curves correspond to the three sublattices of Fig.~\ref{fig:TrimLattice}. In (a), the positions of the cores can be identified by one of the three sublattices having $S^y\!=\!\frac{1}{\sqrt{3}}$, where spins point exactly along one of the four $\langle111\rangle$ axes.\label{fig:Modulationa}} 
\end{figure}

Fig.~\ref{fig:Modulationa} shows the behavior of the $y$-component of the spins, $S^y$, along the $\vec{a}$ direction of the lattice, through two different horizontal cuts of the lattice, one (a) crossing the cores (i.e. going though the edges of the honeycomb superlattice) and another (b) crossing half-way between the cores. The data correspond to the optimal $Z_2$VC obtained for $K/J\!=\!-0.13$ and 276$\times$276 sites, for which the distance between vortex cores is $d\!=\!46$ (see Table~\ref{tab:Z2data}). 

The most remarkable feature of Fig.~\ref{fig:Modulationa} is that $S^y$ shows clear features of abrupt, soliton-like modulations as we cross through the cores (a), while the modulations in (b) are almost harmonic. A similar behavior is shown by the $x$- and $z$-components of the spins along $\vec{c}$ and $\vec{b}$, respectively. This demonstrates the intrinsic non-linear nature of the vortex cores, and is precisely why the LT method cannot describe the ground state by a simple, `3Q' linear superposition of three harmonic waves, one for each cartesian component. Instead, the abrupt behavior can only be recovered by including a very large number of higher harmonics, despite the fact that the amplitudes of higher harmonics in the spin structure factor drop fast, see App.~\ref{app:HarmAmps}.

Incidentally, Fig.~\ref{fig:Modulationa} also shows that the sum of $S^y$ over the three sublattices is almost everywhere equal to zero (consistent with the local 120$^\circ$ structure), except at the immediate vicinity of the cores where the sum is finite. Including the corresponding contributions from $S^x$ and $S^z$ leads to a finite FM canting out of the 120$^\circ$ state at the immediate vicinity of the cores, as discussed above. 
This observation is also the key for understanding the energy competition between Heisenberg and Kitaev exchange. The former is satisfied almost everywhere except near the cores where the total moment is finite. By contrast, the `yy' portion of the Kitaev anisotropy is satisfied only on one-third of the horizontal bonds away from the cores, while near the cores there is a large Kitaev energy gain from all horizontal bonds (similarly for the other types of bonds). So the Kitaev anisotropy is responsible for the spontaneous formation of cores and acts to increase their density, against the Heisenberg exchange.

\subsection{Inequivalent types of $Z_2$ vortices \& their spatial pattern}\label{sec:types} 
There are 24 types of inequivalent $Z_2$ vortices: The three sublattices and four special axes give 12 vortices, and we can get 12 more by time reversal. Each magnetic unit cell encloses all of the 12 former vortices. Their time-reversed versions do not appear in the patterns of Fig.~\ref{fig:111}, because spins do not pass through any of the reverse directions, $[\bar{1}\bar{1}\bar{1}]$, $[\bar{1}11]$,  $[1\bar{1}1]$, and  $[11\bar{1}]$, see also Fig.~\ref{fig:111}~(g).

\subsubsection{Spatial pattern for $K\!<\!0$}
The alternation between inequivalent vortices along different lattice directions follows a very characteristic pattern, according to Figs.~\ref{fig:111} and \ref{fig:Modulationa}. This pattern can be understood by realizing that the Kitaev anisotropy is effective in the regions between vortex domains:  Then the pattern follows from the sign of $K$ (negative in Figs.~\ref{fig:111} and \ref{fig:Modulationa}) and the three-sublattice partition of Fig.~\ref{fig:TrimLattice}. Indeed, the data show that e.g. a `A-$[111]$' vortex gives way to a `C-$[\bar{1}1\bar{1}]$' vortex or a `B-$[\bar{1}1\bar{1}]$' vortex along $\pm\vec{a}$, thus preserving the $y$-component of the domains. Similarly, the $x$- and $z$-components are not altered along $\pm\vec{c}$ and $\pm\vec{b}$ for $K\!<\!0$. 

\subsubsection{Spatial pattern for $K\!>\!0$}
The situation for positive $K$ is analogous, but there are two qualitative differences, both of which can be understood by the form of the long-distance action of the problem around the AF Heisenberg point, see Sec.~\ref{sec:LDT}. First, the direction of the chirality vector is flipped (and so is the sign of its vorticity), which is equivalent with interchanging e.g. sublattices B and C. So, an `A-$[111]$' vortex is now followed by a `B-$[\bar{1}1\bar{1}]$' vortex (and a $C_y\!<\!0$ region) or a `C-$[\bar{1}1\bar{1}]$' vortex (and a $B_y\!<\!0$ region) along $\pm\vec{a}$. Given the three-sublattice partition of Fig.~\ref{fig:TrimLattice}, this alternation corresponds to an AF arrangement of the $y$-component of the domains along $\pm\vec{a}$.

The second difference between the $Z_2$VC's at $K$ and $-K$ is that the characteristic period of modulation, and thus the distance between vortices, is not the same, see Sec.~\ref{sec:dv} below. 

\subsubsection{Threefold symmetry of the  $Z_2$VC state}
From Fig.~\ref{fig:111} we can deduce that the $Z_2$VC state preserves the threefold symmetry axes that pass through the cores of the vortices. Indeed, take for example the center of a type `A-$[111]$' vortex. In real space, the threefold symmetry maps sublattices A to A, and sublattices B to C, while in spin space, it maps the spin components as shown in Eq.~(\ref{eq:sym}). So three successive combined rotations map A-$[\bar{1}\bar{1}1] \mapsto$ A-$[\bar{1}1\bar{1}] \mapsto$ A-$[1\bar{1}\bar{1}]$, and similarly, B-$[\bar{1}\bar{1}1] \mapsto$ C-$[\bar{1}1\bar{1}] \mapsto$ B-$[1\bar{1}\bar{1}]$, which is fully consistent with the patterns of Fig.~\ref{fig:111}.

This symmetry is also reflected in the fact that the three eigenvalues of the spin-inertia tensor $\varmathbb{I}$ discussed above are equal to each other, and the fact that the three spin structure factors have equal magnitudes in all harmonics, see App.~\ref{app:HarmAmps}.

\subsubsection{Superlattice vectors}\label{sec:supervectors}
There are several more things that we learn from Fig.~\ref{fig:111}. First, the vortex superlattice (v), the magnetic superlattice of a given sublattice (s), and the magnetic superlattice of a the full structure (m), are given, respectively, by the translations
\bea
\begin{array}{ll}
\vec{T}_1^{\text{v}}=d ~\vec{a}, & \vec{T}_2^{\text{v}}=d ~\vec{b},\label{eq:T12v} \\
\vec{T}_1^{\text{s}}=2d ~(\vec{a}\!-\!\vec{b}), &~ \vec{T}_2^{\text{s}}=2d ~(\vec{c}\!-\!\vec{b}),\label{eq:T12s}\\
\vec{T}_1^{\text{m}}=2d ~\vec{a}, & \vec{T}_2^{\text{m}}=2d ~\vec{b}, \label{eq:T12m}
\end{array}
\eea 
where $d$ is the distance between vortices (in units of $a$), see below. Each vortex occupies $d^2$ sites, each sublattice FM domain occupies $d^2$ sites (from the given sublattice type), each sublattice unit cell occupies $12d^2$ sites ($4d^2$ from each sublattice type), and each magnetic unit cell of the full structure occupies $4d^2$ sites. Accordingly, the total number of these three types of unit cells are
\be\label{eq:Nmv}
N_{\text{v}}\!=\!\frac{N}{d^2},\quad
N_{\text{s}}\!=\!\frac{N}{12d^2},\quad 
N_{\text{m}}\!=\!\frac{N}{4d^2}. 
\ee
Next, the reciprocal vectors corresponding to $\vec{T}_1^{\text{v}}$ and $\vec{T}_2^{\text{v}}$ are  
\be\label{eq:G12v}
\vec{G}_1^{\text{v}}\!=\!\frac{1}{a}(\frac{2\pi}{d},\!\frac{-2\pi}{\sqrt{3}d}), ~
\vec{G}_2^{\text{v}}\!=\!\frac{1}{a}(0,\!\frac{-4\pi}{\sqrt{3}d}), 
\ee 
which are along the directions of the $\vec{M}$ points, consistent with the dominant wavevector peaks of Fig.~\ref{fig:Ek_H1_Km025}. 
Similarly, the reciprocal vectors corresponding to $\vec{T}_1^{\text{s}}$ and $\vec{T}_2^{\text{s}}$ are  
\be\label{eq:G12s}
\vec{G}_1^{\text{s}}=\frac{1}{a}(\frac{2\pi}{3d},0), \quad
\vec{G}_2^{\text{s}}=\frac{1}{a}(-\frac{\pi}{3d},\frac{\pi}{\sqrt{3}d}), 
\ee
which are along the corners of the BZ. Including the local 120$^\circ$ modulation from all three sublattices shifts these points to $\vec{X}\!+\!\vec{G}_1^{\text{s}}$ and $\vec{X}\!+\!\vec{G}_2^{\text{s}}$ (where $\vec{X}\!=\!\frac{1}{a}(\frac{2\pi}{3},\frac{2\pi}{\sqrt{3}})$ is a corner of the BZ), in agreement with the dominant wavevector peaks of Fig.~\ref{fig:mcmc_spins}. 
The higher harmonics present in the structure factor (see App.~\ref{app:HarmAmps}) are multiples of the fundamental harmonics $\vec{X}\!+\!\vec{G}_{1,2}^{\text{s}}$. In total, for integer $d$, there are at most $4d^2$ harmonics, equal to the number of sites in the magnetic unit cell.

Let us now discuss the distance $d$ between the cores. A first approximation to this number can be obtained from the LT method. Namely, from the distance $q=|\vec{G}_1^{\text{m}}|\!=\!\frac{2\pi}{3a d}$ of the minima $\vec{Q}^{(\gamma)}$ of $\lambda_{\gamma}(\vec{k})$ from the corners of the BZ:
\be\label{eq:d}
d_{\text{LT}}\!=\!\frac{2\pi}{3a q_{\text{LT}}},\quad
a q_{\text{LT}}\!=\!|2\arccos{\frac{1}{2(\frac{K}{J}\!+\!1)}}\!-\!\frac{2\pi}{3}|,
\ee 
where the subscript `LT' indicates that this is the prediction from the LT method. Note that in the vicinity of the AF Heisenberg point, the distance $d_{\text{LT}}$ behaves as
\be\label{eq:dsmallK}
d_{\text{LT}}\!\simeq\!\frac{\pi}{\sqrt{3}} \frac{|J+\frac{7}{6}K|}{|K|},
\ee
suggesting an extra contribution to the effective stiffness coming from the Kitaev anisotropy, that depends on the sign of $K$. We shall come back to this point below.

\subsection{`Optimal' $Z_2$VC's}\label{sec:dv}
We now describe how one can use the knowledge of the detailed structure of the vortex crystals in order to obtain a much more accurate description of $d(K)$. First of all, a finite cluster can accommodate a $Z_2$VC if both the distance between vortices, $d$, and the number of magnetic unit cells per sublattice, $N_{\text{s}}\!=\!\frac{N}{12d^2}$, are integer numbers. With this in mind, we first fix the value of $K/J$ and then we construct states that look as in Fig.~\ref{fig:111}~(a-c), but with the difference that all spins in each domain point strictly along the $\langle111\rangle$ axes. These states can then be used as initial states for a numerical iterative minimization scheme, where we sequentially rotate spins in the direction of their local mean fields, in a random fashion. In most cases, these iterations do not disturb the positions of the vortex cores, nor do they deform the relative domain shapes and sizes, which is important for staying close to very low variational energies. We can repeat this procedure with initial states that have different integer values of $d$, without changing the value of $K/J$, and then choose the crystal with the minimum energy.

\begin{figure}[!t]
\includegraphics[width=0.49\textwidth,clip]{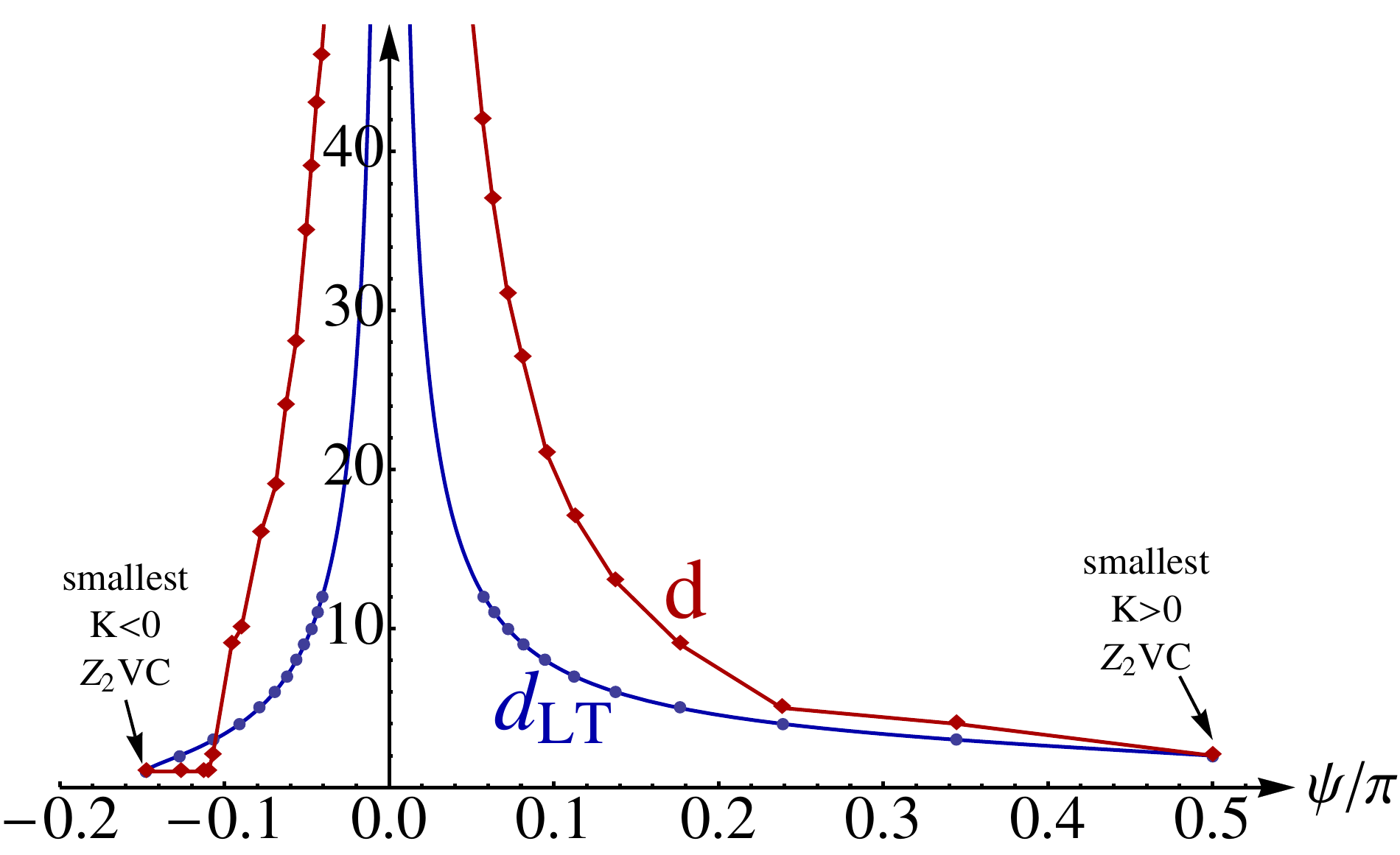}
\caption{Optimal distance $d$ between $Z_2$ vortices as a function of $\psi\!\in\![-\arctan{\frac{1}{2}},\frac{\pi}{2}]$, obtained as described in the text. The most important data associated with these crystals are provided in Table~\ref{tab:Z2data}.  For comparison we also show the approximate prediction $d_{\text{LT}}$ from the LT method, Eq.~(\ref{eq:d}).\label{fig:d_vs_KoJ}} 
\end{figure}

\begin{table}[!t]
\caption{Most important data associated with the optimal $Z_2$VC's found numerically (and shown by red diamond symbols in Fig.~\ref{fig:d_vs_KoJ}) in periodic clusters with spanning vectors $L\vec{a}$ and $L\vec{b}$. Here $E_0/N$ gives the energy per site, $\lambda^\infty_{\text{min}}$ is the low bound given by the LT method (for $L\!=\!\infty$), and $\bs{\kappa}_{\text{min}}$ and $\vec{M}_{\text{max}}$ are the magnitudes of the chirality and magnetization vectors at the cores of the $Z_2$ vortices. The latter two values are not given for $K/J\!=\!1.88$ because the corresponding crystal could not be fully optimized.\label{tab:Z2data}}
\begin{ruledtabular}
\begin{tabular}{@{}ccccccccc@{}} 
$K/J$ &  $L$ & $d_{\text{LT}}$& $d$ & $\sqrt{N_{\text{v}}}$ & $E_{0}/N$ & $\lambda^\infty_{\text{min}}$ & $|\bs{\kappa}|_{\text{min}}$ & $|\vec{M}|_{\text{max}}$\\
\hline
-0.13 & 276 &12& 46 & 6 & -1.42466& -1.43266 & 0.873 & 0.841\\
-0.14& 258 & 11& 43 &6& -1.41817 & -1.42747 & 0.878 &0.830\\
-0.15& 234&10&39&6& -1.41159 & -1.42232 & 0.877 &0.832\\ 
-0.16& 210 & 9& 35 &6&  -1.40493& -1.41721& 0.871 &0.843\\
-0.18 & 168 &8& 28 & 6  & -1.39143 & -1.40714& 0.877 & 0.829\\
-0.2& 144& 7&24&6&-1.37773  & -1.39733& 0.873 &0.835 \\
-0.22 & 114 &6& 19 & 6  & -1.36390 & -1.38784& 0.877 & 0.826 \\
-0.25&96 &5 &16&6& -1.34304 & -1.37437& 0.876 &0.824\\ 
-0.29 & 60 &4& 10 & 6  & -1.31564 & -1.35826& 0.876 &0.822\\ 
-0.31 & 54 &3.46& 9 & 6  & -1.30220  & -1.35080 & 0.849 & 0.868\\ 
-0.35 & 12 &3& 2&6 & -1.27916 & -1.33955 
& $\left\{\!\begin{array}{c}0.633\\0.887\end{array}\right.$ 
& $\left\{\!\begin{array}{c}0.934\\0.785\end{array}\right.$ \\
-0.36 & 48  &2.81& 1 & 48 &  -1.27961 & -1.33724& 8/9 & 1\\ 
-0.37 & 48  &2.66& 1 & 48 &  -1.28487 &-1.33519& 8/9 & 1\\
-0.42 & 48  &2& 1 & 48 &  -1.30921 & -1.32956& 8/9 & 1\\ 
-0.5 & $L$  & 1& 1 & $L$ & -1.34164 & -1.34164& 8/9 & 1\\
\hline
0.20& 222&11 &37&6& -1.57226&-1.58527 &0.846&0.847\\ 
0.23& 186 &10 & 31 & 6 &  -1.57843 &-1.59486 &0.855&0.825\\
0.26&162 & 9& 27&6& -1.58345 &-1.60351 &0.847&0.833\\
0.31 & 126  &8& 21 &6 &  -1.58938 & -1.61582&0.834&0.847\\
0.37& 102 & 7 &17& 6& -1.59279 &-1.62716 &0.822&0.858\\ 
0.46 & 78  &6& 13 & 6 &  -1.59134 & -1.63752&0.820&0.834\\ 
0.62 &  54 & 5 & 9 &  6 & -1.57403  & -1.63916&0.783&0.880\\ 
0.93 & 30  &4 &5 & 6 &  -1.51589 & -1.60299&0.683&0.975\\ 
1.88 & 24  &3& 4 &6 & -1.36213 & -1.43402&--&--\\
%--------------------------------------------------------------------------
$\infty$ & $L$  &2& 2 & $L/2$ &  -1 & -1
&$\left\{\!\!\!\begin{array}{c} \sqrt{2/3}~ 8/9 \\ 8/9 \end{array}\right.$ & 1
\end{tabular}
\end{ruledtabular}
\end{table}

The optimal crystals obtained in this way are listed in Table~\ref{tab:Z2data}, and it is clear that their energies are very close to the low-energy bound given by $\lambda_{\text{min}}$, much closer than the energies obtained from unconstrained MC simulations. The optimal crystal solutions are also shown by data points in Fig.~\ref{fig:d_vs_KoJ} (red diamonds), and their positions give our numerical estimate of $d(K)$ (red curve). For comparison we also show the behavior of $d_{\text{LT}}(K)$ (blue curve).  

There is a number of things we learn from this figure. First, the LT method generally underestimates the vortex distance and is a very crude approximation especially near the AF Heisenberg point, where $d/d_{\text{LT}}\!\simeq\!4$ for $|K|/J\!\simeq\!0.1$ (see Table~\ref{tab:Z2data}), and is possibly larger at smaller values of $|K|/J$. This large difference demonstrates once again the strong impact of non-linearities, which are completely missed by the LT method. 

Second, the vortex distance is not the same for $K$ and $-K$. This dependence on the sign of $K$, which is also reflected in the leading behavior of Eq.~(\ref{eq:dsmallK}), originates in an effective exchange anisotropy driven by $K$, see Sec.~\ref{sec:LDT}.

Third, the LT method gives the right answer at the two boundaries of the $Z_2$VC phase, namely when $\psi\!\to\!\left(\frac{\pi}{2}\right)^-$ and $\psi\!\to\!\left(-\arctan{\frac{1}{2}}\right)^+$, where, $d\!=\!2$ and $d\!=\!1$, respectively. This happens because at these boundaries the minima of $\lambda_\gamma(\vec{k})$ correspond to commensurate momenta, and as such they deliver states that satisfy the spin length constraints, see Sec.~\ref{sec:SmallestVortices}. 

\begin{figure*}[!t]
\includegraphics[width=0.48\textwidth,clip]{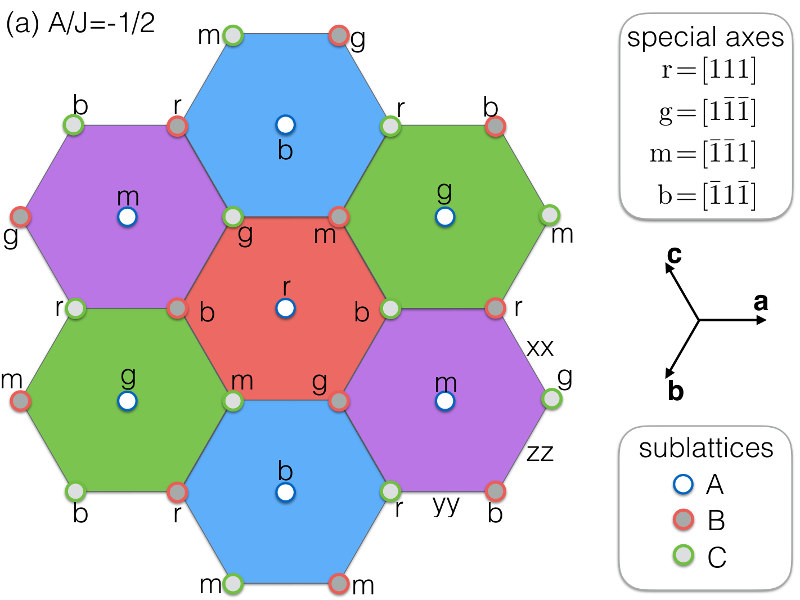}
\includegraphics[width=0.35\textwidth,clip]{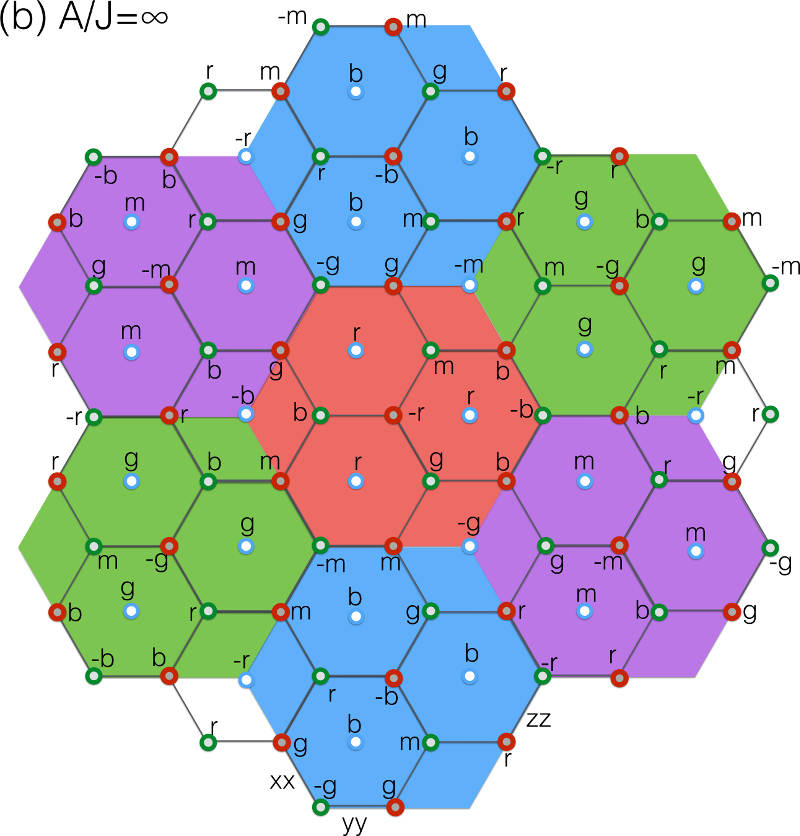}
\caption{The two `smallest vortex crystals' corresponding to the commensurate LT solutions at $\psi\!=\!-\arctan{1/2}$ (a) and $\pi/2$ (b), see text. The colors show the `smallest vortex domains' in the A-sublattice, which enclose one (a) or three sites (b).  \label{fig:SmallestVortex}} 
\end{figure*}

\subsection{The `smallest' vortex crystals}\label{sec:SmallestVortices} 
The above commensurate states were discussed briefly in Secs.~\ref{sec:kitaev} and \ref{sec:FM}, and are shown in Fig.~\ref{fig:SmallestVortex}. They have all the characteristic features of the vortex crystal states that we found numerically deep inside the $Z_2$VC phase, and as such they can be thought of as the $Z_2$VC states with the smallest possible sublattice FM domains.

Specifically, at $K/J\!=\!-1/2$ [Fig.~\ref{fig:SmallestVortex}(a)], each spin points along one of the four $\langle111\rangle$ axes, and can be considered as the core of a `discrete' $Z_2$ vortex. The corresponding sublattice FM domain consists of this single spin only. The six spins surrounding the core rotate in such a way that a total vorticity of $2\pi$ is recovered in discrete steps. The symmetry and spatial pattern of the vortices is exactly the same with the states found numerically away from this point. Furthermore, the magnitudes of the chirality vector and the total moment are, respectively, $|\bs{\kappa}|\!=\!\frac{8}{9}$ and $|\vec{M}|\!=\!1$, the same for all elementary triangles. These values are very close to the corresponding numerical values (minimum chirality $|\bs{\kappa}|_{\text{min}}$ and maximum local moment $|\vec{M}|_{\text{max}}$, appearing at the cores), in all crystal states found in the entire stability region of the $Z_2$VC phase, see Table~\ref{tab:Z2data}. 

For $K/J\!=\!\infty$ [Fig.~\ref{fig:SmallestVortex}(b)] we have $d\!=\!2$ and each sublattice FM domain consists of four sites (of the given sublattice). Take for example the central shaded (red) hexagon of Fig.~\ref{fig:SmallestVortex}(b), which corresponds to an `A-$[111]$' domain. The three A-sites in the interior of this domain point along $[111]$ (denoted by `r'). The fourth spin is shared with the neighboring domains, with directions along $[11\bar{1}]$ (`-m'), $[1\bar{1}1]$ (`-b'), and $[\bar{1}11]$ (`-g'), but gives a total moment along $[111]$ as well. So each sublattice domain has a large moment along one of the $\langle111\rangle$ axes, and the $Z_2$ vortices can be seen by looking at the rotation of the directions of the B- and C-sublattice domains. In terms of elementary triangles, this structure has two inequivalent types. Both of them have $|\vec{M}|\!=\!1$, but $|\bs{\kappa}|\!=\!\frac{8}{9}$ in 1/3 of the triangles and $|\bs{\kappa}|\!=\!\frac{\sqrt{2}}{\sqrt{3}}\frac{8}{9}$ in the remaining 2/3 of the triangles. A similar structure with tho types of elementary triangles is shared by the $d\!=\!2$ state found numerically at $K/J\!=\!-0.35$, see Table~\ref{tab:Z2data}.

Finally, Fig.~\ref{fig:SmallestVortex} provides an intuitive picture as to what happens as we depart from the boundaries toward the HAF limit $\psi\!=\!0$. The AF Heisenberg interaction tends to align all the spins of a given sublattice into one giant FM domain, and the way this happens is by successively eliminating more and more neighboring cores and enclosing the corresponding spins into the domain, until there is only one domain that can be accommodated by the system, which would correspond to the coplanar 120$^\circ$ state at $\psi\!=\!0$.

%-----------------------------------------------------------------------------------
\section{Physical mechanism: Long-distance theory around the AF Heisenberg point}\label{sec:LDT}
\subsection{Derivation}
Let us now establish the physical ``double-twisting'' mechanism that stabilizes the vortex  phase. The model itself has inversion symmetry, which at first sight would point against the presence of Lifshitz invariants in the low-energy action of the problem. However, inversion symmetry  is spontaneously broken in the 120$^\circ$ state. It is this spontaneous handedness that allows the spin-orbit coupling to generate chiral interactions in the form of Lifshitz invariants. To show this, we follow Ref.~[\onlinecite{DombreRead,Azaria}] and derive the classical action for the long wavelength limit of Eq.~(\ref{eq:ham}) in the vicinity of the Heisenberg point $K\!=\!0$. This is a coarse-grained description which builds on the fact that for very small $K$ each elementary triangle retains a very rigid 120$^\circ$ structure, and thus a local order parameter -- in this case an SO(3) rotation matrix $\varmathbb{R}(\vec{r})$ -- has a well defined meaning. A finite FM canting $\vec{M}(\vec{r})$ out of the 120$^\circ$ structure can also be included in terms of a vector $\vec{L}(\vec{r})$, see below. The derivation of the continuum action then involves rewriting the individual lattice spin degrees of freedom in terms of $\varmathbb{R}(\vec{r})$ and $\vec{L}(\vec{r})$, followed by a Taylor expansion of the energy in the lattice constant $a$.

We begin by fixing the reference 120$^\circ$ state of the pure Heisenberg limit ($K\!=\!0$), around which we wish to expand. To this end we define a fixed reference frame of orthonormal vectors $\vec{e}_1$, $\vec{e}_2$ and $\vec{e}_3\!=\!\vec{e}_1\!\times\!\vec{e}_2$, and write the directions of the three sublattices of the state, labeled by the letters A, B and C (see Fig.~\ref{fig:TrimLattice}), as $\vec{n}_A\!=\!\vec{e}_1$,  $\vec{n}_B\!=\!(-\vec{e}_1\!+\!\sqrt{3}\vec{e}_2)/2$,  and $\vec{n}_C\!=\!(-\vec{e}_1\!-\!\sqrt{3}\vec{e}_2)/2$. 
We then introduce three vector fields $\vec{A}(\vec{r})$, $\vec{B}(\vec{r})$ and $\vec{C}(\vec{r})$, that encode the long distance behavior of the spins in the three sublattices A, B and C (see Fig.~\ref{fig:TrimLattice}), in terms of the spatially dependent $\varmathbb{R}(\vec{r})$ and $\vec{L}(\vec{r})$:
\bea
&&\vec{A}(\vec{r}) \!=\! \varmathbb{R}(\vec{r}) \!\cdot\! \left( \vec{n}_A \!+\! a \vec{L}(\vec{r})\right)\!/\!\left(1\!+\!2a\vec{n}_A\!\cdot\!\vec{L}\!+\!a^2\vec{L}^2\right)^{1/2}\!\!,\nonumber\\
&&\vec{B}(\vec{r}) \!=\! \varmathbb{R}(\vec{r}) \!\cdot\! \left( \vec{n}_B \!+\! a \vec{L}(\vec{r})\right)\!/\!\left(1\!+\!2a\vec{n}_B\!\cdot\!\vec{L}\!+\!a^2\vec{L}^2\right)^{1/2}\!\!,~~~~\\
&&\vec{C}(\vec{r}) \!=\! \varmathbb{R}(\vec{r}) \!\cdot\! \left( \vec{n}_C \!+\! a \vec{L}(\vec{r})\right)\!/\!\left(1\!+\!2a\vec{n}_C\!\cdot\!\vec{L}\!+\!a^2\vec{L}^2\right)^{1/2}\!\!.\nonumber
\eea
The rotation matrix can be parametrized in terms of three mutually orthonormal vector fields as 
\be
\varmathbb{R}(\vec{r})=\left( \bs{\mu}(\vec{r}), \bs{\nu}(\vec{r}),\bs{\pi}(\vec{r})\right),
\ee
where the column vectors $\bs{\mu}(\vec{r})\!=\!\varmathbb{R}(\vec{r})\!\cdot\!\vec{e}_1$, $\bs{\nu}(\vec{r})\!=\!\varmathbb{R}(\vec{r})\!\cdot\!\vec{e}_2$, and $\bs{\pi}(\vec{r})\!=\!\varmathbb{R}(\vec{r})\!\cdot\!\vec{e}_3\!=\!\bs{\mu}(\vec{r})\!\times\!\bs{\nu}(\vec{r})$. The vector $\vec{L}$ can in turn be parametrized as $L_i\!=\!\vec{L}\!\cdot\!\vec{e}_i$, and the total moment $\vec{M}(\vec{r})\!\simeq\!\vec{A}(\vec{r})\!+\!\vec{B}(\vec{r})\!+\!\vec{C}(\vec{r})$, is then given by 
\be
\vec{M}(\vec{r})=3a \left(\frac{1}{2}L_1 \bs{\mu}+\frac{1}{2}L_2 \bs{\nu}+L_3\bs{\pi}\right)+\mc{O}(a^2)~.
\ee
Next, we perform a Taylor expansion of the fields in $a$ (or, more precisely, in $q a$, where $q$ is the characteristic modulation wavevector) keeping terms up to order $a^2$. Apart from an overall constant, the total energy density $\varepsilon(\vec{r})$, defined as $E\!=\!\int\! \frac{d^2\vec{r}}{a^2}\varepsilon(\vec{r})$, reads: 
\bea
\varepsilon(\vec{r})\!&=&\!
-p_{\text{Lif}}\!\sum_{\bs{\epsilon}}
\left( \mu^{\gamma_{\bs{\epsilon}}}\partial_{\epsilon}\nu^{\gamma_{\bs{\epsilon}}} 
\!-\!\nu^{\gamma_{\bs{\epsilon}}}\partial_{\epsilon}\mu^{\gamma_{\bs{\epsilon}}} \right)
\label{eq:Lif}\\
\!&-&\!\frac{1}{2}p_{\text{Lif}}\!\sum_{\bs{\epsilon}} \Big\{
2L_1 \partial_{\epsilon} (\mu^{\gamma_{\bs{\epsilon}}}\nu^{\gamma_{\bs{\epsilon}}})
\!+\!L_2 \partial_{\epsilon} [(\mu^{\gamma_{\bs{\epsilon}}})^2\!-\!(\nu^{\gamma_{\bs{\epsilon}}})^2]\Big\}
~~~~~~~\label{eq:cc}\\ 
\!&+&\! p_{\text{el-}K}\sum_{\bs{\epsilon}}\!\Big( (\partial_{\epsilon} \mu^{\gamma_{\bs{\epsilon}}})^2 + (\partial_{\epsilon} \nu^{\gamma_{\bs{\epsilon}}})^2 \Big) 
\label{eq:elK}\\
\!&+&\! p_{\text{el-}J}\Big( 
( \partial_{x'} \bs{\mu} )^2 \!+\! ( \partial_{y'} \bs{\mu} )^2 \!+\! ( \partial_{x'} \bs{\nu} )^2 \!+\! ( \partial_{y'} \bs{\nu} )^2
\Big) \label{eq:elJ}~~~~~~~~\\
\!&+&\!p_L\Big(\frac{1}{2}L_1^2+\frac{1}{2}L_2^2+L_3^2\Big), \label{eq:L}
\eea
where the coupling constants 
\be
\begin{array}{c}
p_{\text{Lif}}\!=\!\frac{K a}{2},~
p_{\text{el-}J}\!=\!\frac{\sqrt{3}J a^2}{8},~
p_{\text{el-}K}\!=\!\frac{K a^2}{4\!\sqrt{3}}\\
p_L\!=\!\sqrt{3} (3J\!+\!K)a^2~.
\end{array}
\ee  
The linear derivative terms in (\ref{eq:Lif}) are the Lifshitz invariants that are responsible for the `double twisting' of $\varmathbb{R}(\vec{r})$ and the spontaneous formation of the solitonic cores. The terms in (\ref{eq:elJ}) and (\ref{eq:elK}) are the elastic energy contributions from the Heisenberg and the Kitaev terms respectively. The term in (\ref{eq:L}) gives the energy cost associated with a finite FM canting $\vec{M}(\vec{r})$, while the term in (\ref{eq:cc}) describes a cross-coupling between $\vec{M}(\vec{r})$ and the twisting of $\varmathbb{R}(\vec{r})$.

The solutions corresponding to the above long-distance action can be investigated following the standard Euler-Lagrange method, with appropriate Lagrange multipliers that ensure the orthogonality of $\varmathbb{R}$. Alternatively, an Euler-angle parametrization of $\varmathbb{R}$ may be more appropriate for studying localized vortex solutions. Leaving this non-trivial task for a separate work, we shall focus here on the general qualitative aspects that derive from the form of the action.

%--------------------------------------------------------------------------------------------------
\subsection{Analogy with other systems}
To highlight the analogy to other well-known condensed matter systems with particle-like modulations, let us disregard for the moment the FM canting and the terms $\propto\!p_{\text{el-K}}$ (which can be incorporated into the exchange energy portion by an appropriate redefinition of the metric), and rewrite the remaining terms (\ref{eq:Lif}) and (\ref{eq:elJ}) in a `covariant derivative' form
\be\label{eqn:Etot}
(\ref{eq:Lif})+(\ref{eq:elJ}) \!\to\!\!\!\int\! d^2\vec{r} \sum_{\bs{\epsilon}, ij} \Bigl[ \rho_j \partial_{\bs{\epsilon}} R_{ij}(\vec{r}) 
+ \frac{K}{J}\!\sum_{km} \Gamma_{ij,km}^{\bs{\epsilon}} R_{km}(\vec{r}) \Bigr]^2, 
\ee
where $\{i,j, k,m\}$ run over $\{\vec{e}_1,\vec{e}_2,\vec{e}_3\}$, $\rho_{1,2}\!=\!1$, $\rho_3\!=\!0$, and the antisymmetric fourth-rank tensors $\bs{\Gamma}^{\bs{\epsilon}}$ are given by 
\be\label{eqn:gamma}
\Gamma_{ij,km}^{\bs{\epsilon}}= -\frac{\sqrt{3}}{a}~e_i^{\gamma_{\bs{\epsilon}}}~e_k^{\gamma_{\bs{\epsilon}}}~\varepsilon_{mj}=-\Gamma_{km,ij}^{\bs{\epsilon}}~,
\ee
where $\varepsilon_{mj}\!=\!\delta_{m,1}\delta_{j,2}\!-\!\delta_{m,2}\delta_{j,1}$. The form (\ref{eqn:Etot}) is analogous to the energy of chiral helimagnets~\cite{Dz64}, where the role of the anisotropy is played by the Dzyaloshinskii-Moriya interaction and the role of $Z_2$ vortices is played by skyrmions. The only difference is that the order parameter is there a vector (the magnetization) instead of an SO(3) matrix. Similar quadratic expressions are also known for the other cases mentioned in the introduction, e.g. in cholesteric liquid crystals~\cite{Wright89}. 

Based on the experience with such systems, there are two general classes of potential solutions. The first are helicoidal one-dimensional modulations~\cite{Dz64}, the second are `double-twisted' states which modulate along all possible spatial directions and thus may achieve lower energy. The  $Z_2$VC's belong to the second type of solutions. 

%--------------------------------------------------------------------------------------------------
\subsection{The role of the sign of $K$}
The form of the Lifshitz invariants explains why a sign change of $K$ reverses the sense of rotation of the chirality vector $\bs{\kappa}(\vec{r})$ around the core of a $Z_2$ vortex. This happens because in the Lifshitz terms, the sign change of $K$ can be gauged away by changing $\bs{\nu}\!\mapsto\!-\bs{\nu}$, which amounts to interchanging the sublattices B and C.

Unlike the Lifshitz terms, the sign change of $K$ cannot be gauged away in the exchange anisotropy term (\ref{eq:elK}) generated by $K$. This means that the two opposite signs of $K$ do not give the same distance $d$ between vortices. In particular, a positive (negative) $K$ effectively increases (decreases) the stiffness of the 120$^\circ$ state, leading to $d(|K|)\!>\!d(-|K|)$, which is also  obeyed by the LT approximation $d_{\text{LT}}$, see (\ref{eq:dsmallK}) and Fig.~\ref{fig:d_vs_KoJ}. 

%Some of our numerical $Z_2$ vortex states for positive $K$ were constructed by interchanging B and C in a negative $K$ state, and by adjusting the value of $K$. 

%--------------------------------------------------------------------------------------------------
\subsection{In-plane FM canting}\label{sec:FMcanting}
The cross-coupling term in (\ref{eq:cc}) shows that we should expect a finite FM canting that follows adiabatically the twisting of $\varmathbb{R}(\vec{r})$. This is consistent with our numerical results presented above, which showed a finite canting at the cores of the $Z_2$ vortices. We also know that the smallest $Z_2$VC states discussed in Sec.~\ref{sec:SmallestVortices} have a finite canting everywhere. Of course, the long-distance action written above needs to be supplemented with higher order processes as we go further away from the pure HAF model, but the leading cross-coupling term in (\ref{eq:cc}) already explains that the canting does not arise from a competing instability mechanism, but it is merely a secondary effect that is dragged along by the spontaneous formation of cores.

To analyze this further we integrate out $L_1$ and $L_2$ to obtain 
\be\label{eq:L12}
L_1\!=\!2\xi\!\sum_{\bs{\epsilon}}\! \partial_\epsilon(\mu^{\gamma_{\bs{\epsilon}}}\nu^{\gamma_{\bs{\epsilon}}}),~
L_2\!=\!\xi\!\sum_{\bs{\epsilon}}\! \partial_\epsilon\left( (\mu^{\gamma_{\bs{\epsilon}}})^2\!-\!(\nu^{\gamma_{\bs{\epsilon}}})^2\right),
\ee
where $\xi\!=\!\frac{p_{\text{Lif}}}{p_L}$. Looking back at Fig.~\ref{fig:Modulationa}, for example, and writing, to leading order in $a$, $\bs{\nu}\!\simeq\!\vec{A}(\vec{r})$ and $\bs{\mu}\!\simeq\!-(\vec{A}(\vec{r})\!+\!2\vec{B}(\vec{r})]/\sqrt{3}$, we see that the largest contribution to the right hand sides of (\ref{eq:L12}) arise precisely at the solitonic cores of the $Z_2$ vortices, consistent with what we found numerically.   

However (\ref{eq:L12}) accounts only for the in-plane component of the canting. The out-of-plane component does not couple to the twisting of $\varmathbb{R}(\vec{r})$, and costs a finite amount of energy. Still, as we found numerically, a finite out-of-plane component is also present at the cores of the $Z_2$ vortices, showing that this component is important for sustaining the vortices.    

Finally, we can also look at the feedback effect of the canting on the twisting of $\varmathbb{R}(\vec{r})$. Replacing (\ref{eq:L12}) to the action leads to a renormalized energy density $\varepsilon'(\vec{r})$ in terms of $\varmathbb{R}(\vec{r})$ and $L_3(\vec{r})$ only:
\bea
\varepsilon'(\vec{r})\!&=&\!
-p_{\text{Lif}}\!\sum_{\bs{\epsilon}} 
\left( \mu^{\gamma_{\bs{\epsilon}}}\partial_{\epsilon}\nu^{\gamma_{\bs{\epsilon}}} 
\!-\!\nu^{\gamma_{\bs{\epsilon}}}\partial_{\epsilon}\mu^{\gamma_{\bs{\epsilon}}} \right)
\label{eq:Lif2}\\
\!&-&\! \frac{p_{\text{Lif}}^2}{p_L} \!\sum_{\bs{\epsilon}}\Big\{ 
\left( \partial_{\epsilon} (\mu^{\gamma_{\bs{\epsilon}}}\nu^{\gamma_{\bs{\epsilon}}}) \right)^2
\!+\!\frac{1}{4} \left( \partial_{\epsilon} [(\mu^{\gamma_{\bs{\epsilon}}})^2\!-\!(\nu^{\gamma_{\bs{\epsilon}}})^2]\right)^2\Big\}
~~~~~~~\label{eq:cc2}\\ 
\!&+&\! p_{\text{el-}K}\sum_{\bs{\epsilon}}\!\Big( (\partial_{\epsilon} \mu^{\gamma_{\bs{\epsilon}}})^2 + (\partial_{\epsilon} \nu^{\gamma_{\bs{\epsilon}}})^2
\Big)\label{eq:elK2}\\
\!&+&\! p_{\text{el-}J}\Big( 
( \partial_{x'} \bs{\mu} )^2 \!+\! ( \partial_{y'} \bs{\mu} )^2 \!+\! ( \partial_{x'} \bs{\nu} )^2 \!+\! ( \partial_{y'} \bs{\nu} )^2
\Big) \label{eq:elJ2}~~~~~~~~\\
\!&+&\!p_L L_3^2.\label{eq:L2}
\eea
The new quadratic derivative term that appears in (\ref{eq:cc2}) gives an extra contribution to the effective exchange anisotropy, which is now quartic in the elements of $\varmathbb{R}$.

\section{Discussion}\label{sec:disc}
The fact that the cores of the $Z_2$ vortices are pinned at the lattice sites by the Kitaev anisotropy suggests that the whole stability region of the $Z_2$VC phase may consist of a cascade of transitions between commensurate vortex crystals, where the vortex distance is an integer multiple of the lattice spacing. This may be true especially in the regions close to the commensurate boundaries of the phase (where the Kitaev anisotropy is large enough), which in turn would correspond to an (incomplete) Devil's staircase scenario.

Next, we discuss the stability region of the $Z_2$VC phase. The region shown in Fig.~\ref{fig:phases} concerns the situation at zero temperature and for classical spins only. The effect of thermal or quantum fluctuations can be gauged by examining the effect of fluctuations in the neighboring phases, which was covered already in Secs.~\ref{sec:KitaevThermal}, \ref{sec:KitaevQuantum}, and \ref{sec:FMfluctuations}. According to that  discussion, thermal and quantum fluctuations favor spins pointing along the cubic axes and not along the $\langle111\rangle$ axes, so the two smallest $Z_2$VC's at the two boundaries of the $Z_2$VC region are penalized. This means that the stability region of the $Z_2$VC phase should shrink by fluctuations, provided the corresponding effect of fluctuations of the $Z_2$VC phases themselves is weak enough (close to the boundaries). This scenario is confirmed by the numerical results of Becker {\it et al}~\cite{trebst14}, for the boundary with the AF Kitaev phase and for $S=1/2$ spins.

Let us now turn to possible realizations. Naturally, the $Z_2$ vortex phase can be observed on triangular systems with AF $J\!>\!0$ and small $|K|\!\ll\!J$. By virtue of the local four-sublattice rotation of Sec.~\ref{sec:model}, this regime can be rigorously mapped into the FM regime $J'\!=\!-J\!<\!0$ and $K'\!=\!K+2J\!\approx\!2J$, thus extending the applicability of the present findings to FM coupled systems as well. One particular family of compounds closely related to this case is the undoped triangular cobaltates CoO2, which feature $\sim\!90^\circ$ O-Co-O bond angles~\cite{Khaliullin05}. The more recently discovered triangular-lattice iridate Ba$_3$IrTi$_2$O$_9$~\cite{Dey:2012dk} seems to fall into the first, AF category, as it features a strong AF $J$ and small anisotropy $|K|\!\ll\!J$, see also discussion by Becker {\it et al.}~\cite{trebst14}
However, this compound has an acentric crystal structure by itself, and therefore any conventional AF order would be already twisted by Dzyaloshinskii-Moriya interactions, provided further anisotropies do not suppress this possibility. Hence, in this material one has to distinguish between such a more conventional AF Dzyaloshinskii spiral state and the present mechanism of the Heisenberg-Kitaev model, which exists also in any centrosymmetric triangular lattice compound.

Another family of correlated spin-orbital compounds with the required hexagonal symmetry of the Kitaev anisotropy are the honeycomb systems with nearly 90$^\circ$ bond angles, such as the iridates Na$_2$IrO$_3$ and Li$_2$IrO$_3$, for which the Kitaev Hamiltonian was originally discussed~\cite{Jackeli:2009p2016,Chaloupka:2013ju,Chaloupka:2010p2461}. While these prototypical honeycomb compounds do not manifest such physics~\cite{footnoteNaLi213}, an extensive body of results, both from experiment~\cite{Ir213_choi_2012,Ir213_ye_2012, Ir213_zigzag_liu_2011,Ir213_jkj2j3_singh_2012,Ir213_LivsNa_cao_13} and first-principles {\it ab initio} calculations~\cite{QCNa213,QCLi213}, has consistently revealed that such honeycomb systems are inherently frustrated, due to the presence of second (and possibly third) neighbor exchange interactions $J_2$ ($J_3$). This opens the possibility for realizing particle-like modulations provided the frustration is strong enough to suppress the more conventional collinear phases of honeycomb magnets~\cite{Fouet2001} and spontaneously break the inversion symmetry. In the light of our results for the triangular lattice, this scenario becomes particularly evident in the limit $J_2\!\gg\!J$ where the honeycomb lattice decomposes into two nearly decoupled triangular sublattices, each one showing a nearly 120$^\circ$ order. The spin-orbit interactions will then again generate Lifshitz invariants along several spatial directions leading to particle-like modulated phases. Preliminary MC simulations indicate that the defect lattice indeed survives coupling of the sublattices by moderate $J$.

Summarizing, we have presented a generic mechanism for the condensation of particle-like magnetization modulations in correlated spin-orbital coupled hexagonal systems, like the triangular and frustrated honeycomb iridates. In analogy to the experiments on chiral ferromagnetic helimagnets, magnetic small-angle neutron~\cite{muehlbauer2009} or X-ray~\cite{Langner2014} scattering  methods can be used to study such extended mesophases~\cite{muehlbauer2009}, as diffraction is also able to detect the chiral long-period modulation of a primarily AF order~\cite{Zheludev1997}. Direct microscopic observation of AF textures by magnetic imaging is presently very difficult, similarly to the imaging of AF domain states. However, the $Z_2$-vortex lattice is a defect-ordered state which implies a strong inhomogeneity of the magnetic moments near the singular defect cores. Measurements of static internal hyperfine field distributions, e.g., by NMR, $\mu$SR or M\"ossbauer methods, should be able to discern a $Z_2$-vortex lattice ground state from other unconventional magnetic orders like spin-liquids, helimagnetic or skyrmionic ground states that are essentially homogeneous and free of defects.

\acknowledgments
This research was sponsored by the Deutsche Forschungsgemeinschaft (DFG) under the Emmy-Noether program. We thank G. Khaliullin, N. B. Perkins and P. Orth for helpful discussions.

\section{Appendix}

\subsection{Harmonic amplitudes}\label{app:HarmAmps}
\begin{figure}[!t]
\includegraphics[width=0.99\columnwidth,trim= 0 0 0 0,clip]{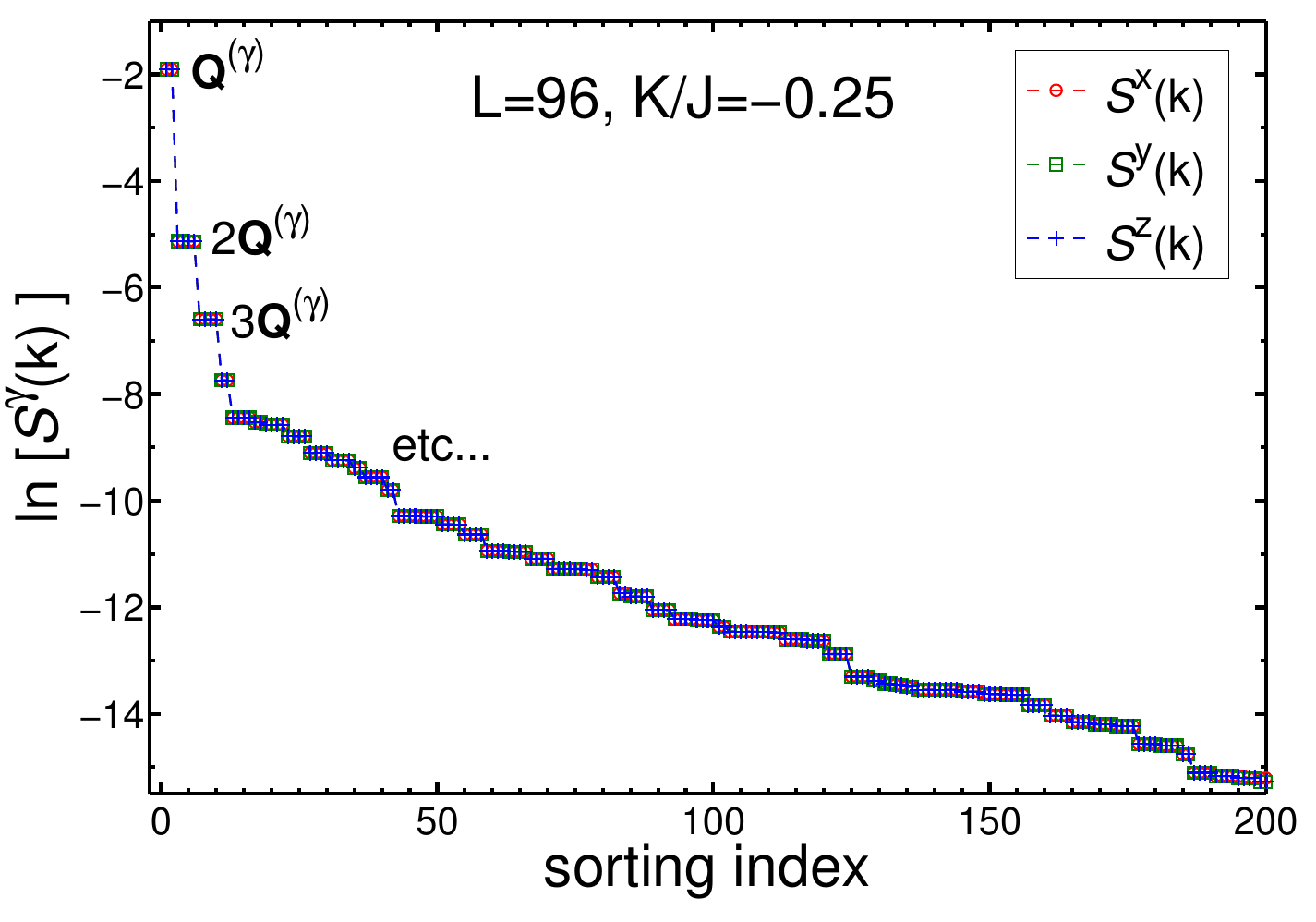}
\caption{Logarithm of the amplitudes $\mc{S}^{\gamma}(\vec{k})$ of the spin structure factor for all higher harmonics of $\vec{Q}^\gamma$, that appear in the optimal $Z_2$VC ground state for 96$\times$96 sites and $K/J\!=\!-0.25$, for which $d\!=\!16$. Only the first 200 momenta are shown for which $\mc{S}^{\gamma}(\vec{k})\gtrsim 10^{-15}$.\label{fig:HarmAmps}} 
\end{figure}

Figure \ref{fig:HarmAmps} shows the amplitudes $\mc{S}^{\gamma}(\vec{k})$ of the spin structure factor of the optimal $Z_2$VC ground state for 96$\times$96 sites and $K/J\!=\!-0.25$, in terms of a sorting index. Only the first 200 momenta are shown, for which  $\mc{S}^{\gamma}(\vec{k})\gtrsim 10^{-15}$. The different spin components $\gamma$ give essentially identical results down to this precision. The maximum amplitudes correspond to the first harmonics $\pm\vec{Q}^{(\gamma)}$ (which coincide with the momenta $\vec{X}\!+\!\vec{G}_1^{\text{s}}$ and $\vec{X}\!+\!\vec{G}_2^{\text{s}}$ of Sec.~\ref{sec:supervectors}) and are equal to $\mc{S}^{\gamma}(\pm\vec{Q}^{(\gamma)})\!\simeq\!0.149$. The second largest amplitudes correspond to four points in the BZ, which are related to the second harmonics $\pm2\vec{Q}^{(\gamma)}$ by reciprocal vectors, with $\mc{S}^{\gamma}(\pm2\vec{Q}^{(\gamma)})\!\simeq\!0.0059$. The next amplitudes are $\mc{S}^{\gamma}(\pm3\vec{Q}^{(\gamma)})\!\simeq\!0.00136$, etc. So the amplitudes drop very quickly as we go to higher and higher harmonics. Note that, in total, there are at most $4d^2$ harmonics, equal to the number of sites in the magnetic unit cell.

The above numbers can also give an idea about the role of the higher harmonics in satisfying the spin length constraints. The latter correspond to the equations:
\be
\sum_{\vec{k}} \vec{S}_{\vec{k}}\cdot \vec{S}_{\vec{q}-\vec{k}} = S^2 \delta_{\vec{q},0}~,
\ee
where here $S^2\!=\!1$. To see the role of the harmonics in the spin length, it suffices to take $\vec{q}=0$ above and then calculate the contribution to the left hand side over restricted sums of momenta $\vec{k}$ that include up to a certain number of harmonics. For instance, if we include only the first harmonics we get 
\be
\sum_{\vec{k}\in \{\pm\vec{Q}^{(\gamma)}\}} \!\!\vec{S}_{\vec{k}}\cdot \vec{S}_{-\vec{k}} \!=\! 6 \mc{S}^{x}(\vec{Q}^{(x)}) \!=\! 0.894.
\ee 
This means that including higher harmonics should give the remaining 11\% of the spin length.  Including the second harmonics gives:
\be
\sum_{\vec{k}\in \{\pm\vec{Q}^{(\gamma)},\pm2\vec{Q}^{(\gamma)}\}} \!\!\vec{S}_{\vec{k}}\cdot \vec{S}_{-\vec{k}} \!=\! 6 \mc{S}^{x}(\vec{Q}^{(x)})\!+\!12 \mc{S}^{x}(2\vec{Q}^{(x)}) \!=\! 0.9648,
\ee 
and so on. The full spin length is recovered asymptotically by including all of the $4d^2$ harmonics. 

The higher harmonics are also important for the soliton-like modulations of the spins in the close vicinity of the cores of the $Z_2$ vortices (Fig.~\ref{fig:Modulationa}~(a)).

\subsection{Monte-Carlo Simulations and finding the vortices} 
The classical Monte-Carlo data presented in Figs.~\ref{fig:nematic}, \ref{fig:mcmc_spins} and \ref{fig:kappa_vort} were obtained using simulated annealing down to low temperatures and lattices of up to $96\times 96$ sites. We complemented the Monte-Carlo simulations by numeric optimization starting from a low-temperature configuration. The data shown here were obtained by simulated annealing down to $\beta J =100$, followed by optimization, for $72\times 72$ sites. Low-temperature snapshots give similar results. 

The orientation of the plane containing the local $120^\circ$ order is given by the vector chirality $\bs{\kappa}(\vec{r})$ defined in Eq.~(\ref{eq:kappa}) from the three spins around upwards pointing triangles. In the $120^\circ$ states, $|\bs{\kappa}({\bf r})|=1$ and it points out of the plane of the spins. The rotation of $\bs{\kappa}(\vec{r})$ is calculated along closed loops on the dual lattice given by every third upward pointing triangle. The results shown here were obtained by rhombus-loops connecting four triangles at ${\bf r}$, ${\bf r}+3{\bf a}$ , ${\bf r}+{\bf a}-{\bf b}$ and ${\bf r}+{\bf a}-{\bf c}$, results with other rhombus loops and with triangular loops connecting only three triangles are equivalent. If all rotations along a loop were around the same axis, the vorticity would be quantized to 0 or 1 on each individual plaquette. As the rotation axes can vary, this does not have to hold. In some places, it adds to one on two neighboring plaquettes. 

\bibliographystyle{apsrev4-1}
%\bibliographystyle{aipnum4-1}
%\bibliography{compass,iridates,frustration,misc,vortices}
%\end{document}

\end{document}